%% file: 34680corr.tex
\documentclass[traditabstract]{aa} 
\usepackage{natbib} 
\usepackage[varg]{txfonts} 
\usepackage{xcolor}             
\usepackage{verbatim}     
\usepackage{xtab,afterpage}  
\usepackage{graphicx} 
\usepackage{tikz} 
\usepackage{epsfig} 
\usepackage{subfig} 
\usepackage{longtable} 
 
\newcommand{\cai}{\ion{Ca}{i}} 
\newcommand{\caii}{\ion{Ca}{ii}} 
\newcommand{\caiii}{\ion{Ca}{iii}} 
\newcommand{\teff}{$T_\mathrm{eff}$} 
\newcommand{\logg}{$\log\,g$} 
\newcommand{\feh}{\mbox{[Fe/H]}} 
\newcommand{\fig}[1]{Fig. \ref{#1}} 
\newcommand{\taufive}{$\tau_{5000}$} 
\newcommand{\acorr}{$\Delta\rm{A(Ca)}_{_{\rm{NLTE-LTE}}}$} 
\newcommand{\weq}{w$_{eq}$} 
\newcommand{\Aca}[2]{A(Ca{#2})$_{_{\rm{#1}}}$}

\begin{document} 
 
\title{Ca line formation in late-type stellar atmospheres: I. The model atom} 
\author{ 
Y. Osorio\inst{1,2} \and K. Lind\inst{3,4} \and P. S. Barklem\inst{3} \and C. Allende Prieto\inst{1,2} \and Oleg Zatsarinny\inst{5}  
} 
\institute{Instituto de Astrof\'isica de Canarias, E-38205 La Laguna, Tenerife, Spain \and Departamento de Astrof\'isica, Universidad de La Laguna (ULL), E-38206 La Laguna, Tenerife, Spain \and Theoretical Astrophysics, Department of Physics and Astronomy, Uppsala University, Box 516, SE-751 20 Uppsala, Sweden \and Max Plank Institute for Astronomy, K\"onigstuhl 17, D-69117 Heidelberg, Germany \and Department of Physics and Astronomy, Drake University, Des Moines, IA 50311, USA} 
\offprints{Yeisson Osorio} 
\mail{yeisson.osorio@iac.es} 
\titlerunning{NLTE Ca line formation in late-type stellar atmospheres.} 
\authorrunning{Osorio Y. et al.} 
\date{} 
\abstract 
   {Departures from local thermodynamic equilibrium (LTE) distort the calcium abundance derived from stellar spectra in various ways, depending on the lines used and the stellar atmospheric parameters. The collection of atomic data adopted in non-LTE (NLTE) calculations must be sufficiently complete and accurate.} 
   {We derive NLTE abundances from high-quality observations and reliable stellar parameters using a model atom built afresh for this work, and check the consistency of our results over a wide wavelength range with transitions of atomic and singly ionised calcium.} 
   {We built and tested \cai\ and \caii\ model atoms with state-of-the-art radiative and collisional data, and tested their performance deriving the Ca abundance in three benchmark stars: Procyon, the Sun, and Arcturus. We have excellent-quality observations and accurate stellar parameters for these stars. Two methods to derive the LTE / NLTE abundances were used and compared. The LTE / NLTE centre-to-limb variation (CLV) of Ca lines in the Sun was also investigated.} 
   {The two methods used give similar results in all three stars. Several discrepancies found in LTE do not appear in our NLTE results; in particular the agreement between abundances in the visual and infra-red (IR) and the \cai\ and \caii\ ionisation balance is improved overall, although substantial line-to-line scatter remains. The CLV of the calcium lines around 6165~\AA\ can be partially reproduced. We suspect differences between our modelling and CLV results are due to inhomogeneities in the atmosphere that require 3D modelling.} 
  {} 
\maketitle

\section{Introduction} 
 
Calcium produced in supernovae explosions is one of the most useful elements for the quantitative analysis of stellar spectra, since it produces numerous absorption lines, many of which have accurate oscillator strengths measured in the laboratory \citep{S,SG,SR}. The \caii\ IR triplet (8\,500-8\,600 \AA) is one of the most extensively used spectral features to study radial velocities and metallicities of distant stars and stellar systems \citep{1989Ap&SS.157...15T,2011A&A...535A.106K}, and is the target of the Radial Velocity Spectrometer on board the Gaia mission \citep[Sect. 3.3.7]{GAIAmission}.  The \caii\ $H$ and $K$ resonance lines at $\sim$3\,950 \AA, on the other hand, are essentially the only metallicity indicator accessible at intermediate or low spectral resolution in extremely metal-poor stars \citep{2004ApJ...603..708C}, in addition to being an invaluable tool to probe the interstellar medium \citep{2010A&A...510A..54W} or solar and stellar chromospheres \citep{Hall2008,2015A&A...577A...7S}.  
 
The separation of the [Ca/Fe] abundance ratios found in thin/thick-disc star sequences with metallicity is smaller compared to other $\alpha$ elements \citep[see Fig. 14 of ][]{2006MNRAS.367.1329R}; a more precise and accurate determination of the calcium abundances is required in order to make this separation more clear.   
Calcium also has observable lines along a broad range of wavelengths in late-type stellar spectra. To ensure the consistency of the derived abundances for transitions in different parts of the spectrum is important in order to make reliable comparisons among different surveys. 
The Apache Point Observatory Galactic Evolution Experiment \citep[APOGEE,][]{Majewski_2017} is part of the Sloan Digital Sky Survey \citep[SDSS-III,][]{Eisenstein_2011} and is observing stars in the northern hemisphere in the H-band (15\,090 - 16\,990 \AA). Its successor  APOGEE-2, part of SDSS-IV \cite{2017AJ....154...28B}, extends the sample to the southern hemisphere. The MaSTAR program (MANGA stellar library, part of SDSS-IV) will also observe northern stars in the optical (3\,600-10\,350 \AA) while the GALAH Survey \citep{2014IAUS..298..322A} observes stars in the southern hemisphere at optical wavelengths (4\,700 - 7\,890 \AA); in those surveys Ca lines are observed for hundreds of thousands of stars.

Early work on NLTE effects for calcium performed in the Sun and Procyon indicated that small corrections should be expected for solar-type stars \citep{1987A&A...181..103S}, but later work has shown that this situation changes depending on the stellar spectral type and luminosity.

NLTE studies on Ca have been done since the 1970s, starting with the ground-breaking work by \cite{1970PASP...82..169L} on \caii\ lines in the Sun. Up to the mid 1980s, NLTE studies on Ca  focused on the $H$~and~$K$ \caii\ lines \citep{1971SoPh...20....3M,1976ApJ...205..165A,1985SoPh...98....1S}.  \cite{1985A&A...149...21W} compared their NLTE calculations with observations of the Sun and Procyon, and found that Ca is more sensitive to NLTE in the latter star. \cite{1991MNRAS.251..369D} calculated the first grid of NLTE calculations, spanning 5400<\teff<6200, 1.0<\logg<4.5, and  -1<\feh<0, and finding enhanced NLTE effects at low \logg\ values and found that departures from LTE in giants can significantly vary among different multiplets of \cai\ lines. An important conclusion from these studies is that the thermalisation of the Ca populations in late-type stars is sensitive to inelastic collisions with hydrogen, and the over-ionisation of \cai\ by UV photons. 
 
\cite{2000ApJ...541..207I}, in the first large-scale application of NLTE calculations for Ca, collected equivalent widths (\weq) from the literature and computed NLTE differential abundances for some 250 dwarf/sub-giant stars, studying the implications for Ca and Mg in Galactic evolution.

More recently, \cite{2007A&A...461..261M} performed a detailed NLTE study of Ca in cool stars with a comprehensive model atom with 63 \cai\ levels, 37 \caii\ levels, and the ground level of \caiii. Energy levels were adopted from the NIST atomic spectra database. Bound-bound and bound-free radiative data were obtained from the Opacity Project available through the TOPBASE database \citep{TOPBASE}. Hydrogen collisions for \cai\ were included via empirically scaled rates from the Drawin formula \citep{1968ZPhy..211..404D,1984A&A...130..319S}. Regarding electron collisions, excitation rates from the ground level of \cai\ were taken  from \cite{2001ADNDT..77...87S}, and for \caii, for transitions involving the seven lowest levels, from \cite{1983MNRAS.203.1269B}. For the remaining allowed and forbidden transitions the impact parameter method (IPM) from \cite{1962amp..conf..375S} and a thermally averaged collision strength of 1.0 \citep{1976asqu.book.....A} were adopted, respectively. Electron collisional ionisation using the Seaton formula \citep{1962amp..conf..375S} was also implemented in this study. Almost all later work for calcium Ca in NLTE until now used modified versions of this model atom.

\cite{2012A&A...541A.143S} used the same model atom from \cite{2007A&A...461..261M}, except that they substituted the IPM for electron collisional excitation with the \cite{1962ApJ...136..906V} formula. They derived calcium abundances for lines in the spectrum of Procyon, finding good agreement between the results for lines of neutral and singly ionised calcium: A(Ca)=6.25$\pm$0.04 and 6.27$\pm$0.06 dex from \cai\ and \caii\ lines respectively, in contrast with the values derived by \cite{2007A&A...461..261M}: 6.19$\pm$0.04 and 6.38$\pm$0.05 dex from \cai\ and \caii\ lines, respectively.

An important conclusion from the \cite{2007A&A...461..261M} study is the sensitivity of the derived \ion{Ca}{i/ii} abundance ratio to inelastic hydrogen collisions. Recently, \cite{2017A&A...605A..53M} used an upgraded version of their model atom from 2007. This update included inelastic calcium-hydrogen collisions from \cite{2016A&A...587A.114B}, avoiding the need for adopting the Drawin formula, and calcium-electron collisions for \caii\ from \cite{2007A&A...469.1203M}. Their results for Procyon were  6.26$\pm$0.05 and 6.32$\pm0.05$ dex from \cai\ and \caii\ lines, respectively, bringing the two average values closer than in their earlier study.

In this paper, we assembled from scratch new model atoms for neutral and singly ionised calcium suitable for the determination of Ca abundance\footnote{Calcium abundance in this work has the usual definition: A(Ca)=$\log(N_{\rm{Ca}}/N_{\rm{H}})+12$, where $N_{\rm{X}}$ is the number density of element X.} in the visual and in the IR where many of the lines come from high excitation levels. We decided to perform this study to take advantage of recently published \mbox{H-Ca} inelastic collisional rates, electron collisional excitation rates calculated via the R-matrix method, new theoretical calculations of oscillator strengths, and, for the first time, updated  methods for the treatment of hydrogen collisional excitation rates involving Rydberg levels of \cai\ and electron collisional ionisation rates for \cai. Our model atom is publicly available\footnote{The model atom used in this work will be available in A\&A, and also will be part of a collection of model atoms for several species that will be presented in a forthcoming paper.}. 
With the new model atoms, we revisit the formation of calcium lines in the photospheres of the Sun, Procyon, and Arcturus ($\alpha$~Boo). We confront our calculations with ultra-high-quality observations for these stars in both the optical and the IR, comparing the inferred abundances. In order to study NLTE effects over such a wide wavelength range, we decided to focus this work on stars with reliable parameters and excellent observations; the interesting case of metal-poor stars will be treated in a forthcoming paper. In Section 2 we describe our new model atoms, while Section 3 addresses the adopted observations for the reference stars. Section 4 is devoted to the statistical equilibrium calculations, and Section 5 examines the impact of departures from LTE on the interpretation of the spectra of the reference stars. Section 6 summarises our main conclusions.

\section{Model atom} 
 
\begin{figure*}[!ht] 
\begin{tabular}{r@{\hspace{-0.1\textwidth}}l} 
\hspace{-0.08\textwidth}\begin{tikzpicture} 
\node[anchor=south east, inner sep=0] (image) at (0,0) {  
\subfloat{\includegraphics[height=0.39\textwidth]{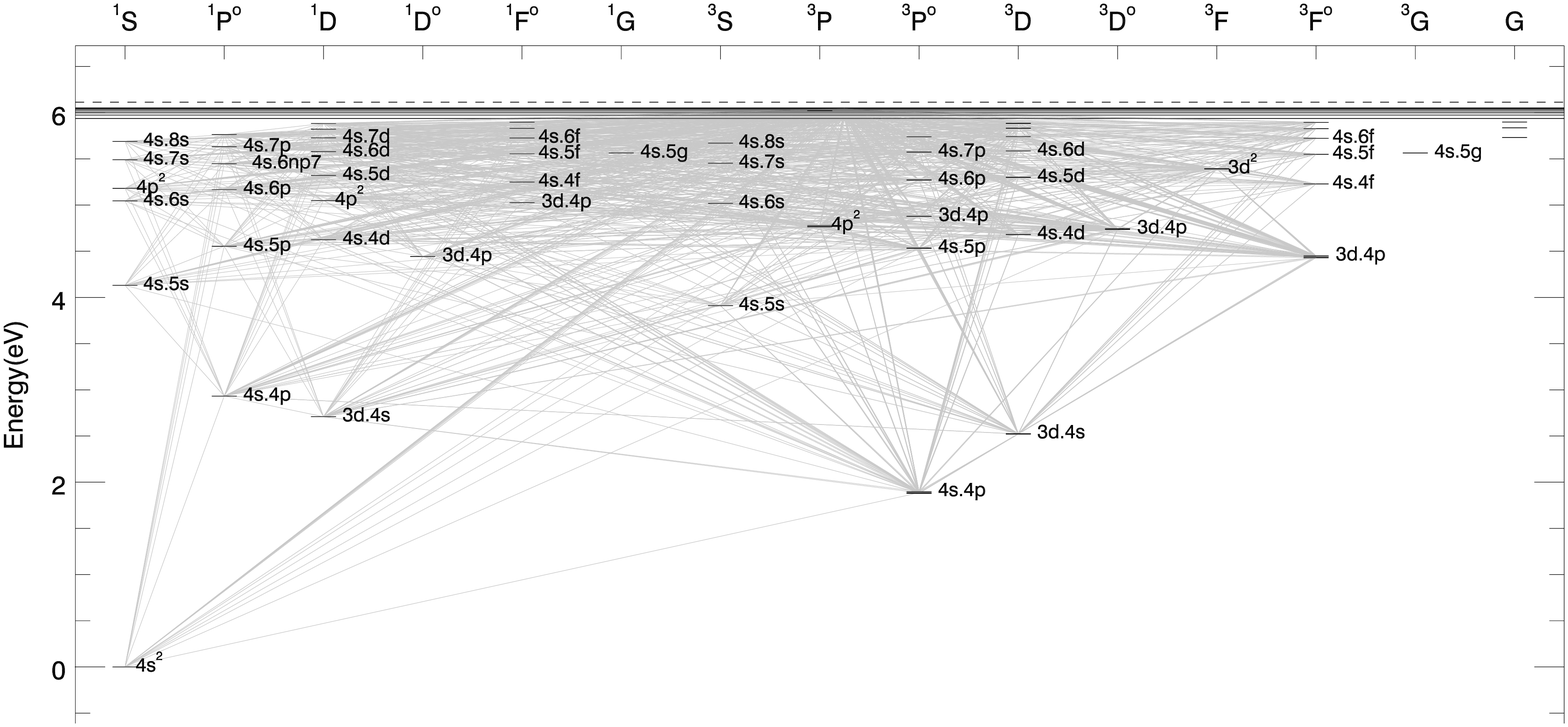}} 
}; 
\node at (-0.075\textwidth,0.1\textwidth) {\scalebox{1.5}{\cai}};   
\end{tikzpicture} 
&  
\begin{tikzpicture} 
\node[anchor=south east, inner sep=0] (image) at (0,0) {  
\subfloat{\includegraphics[height=0.39\textwidth]{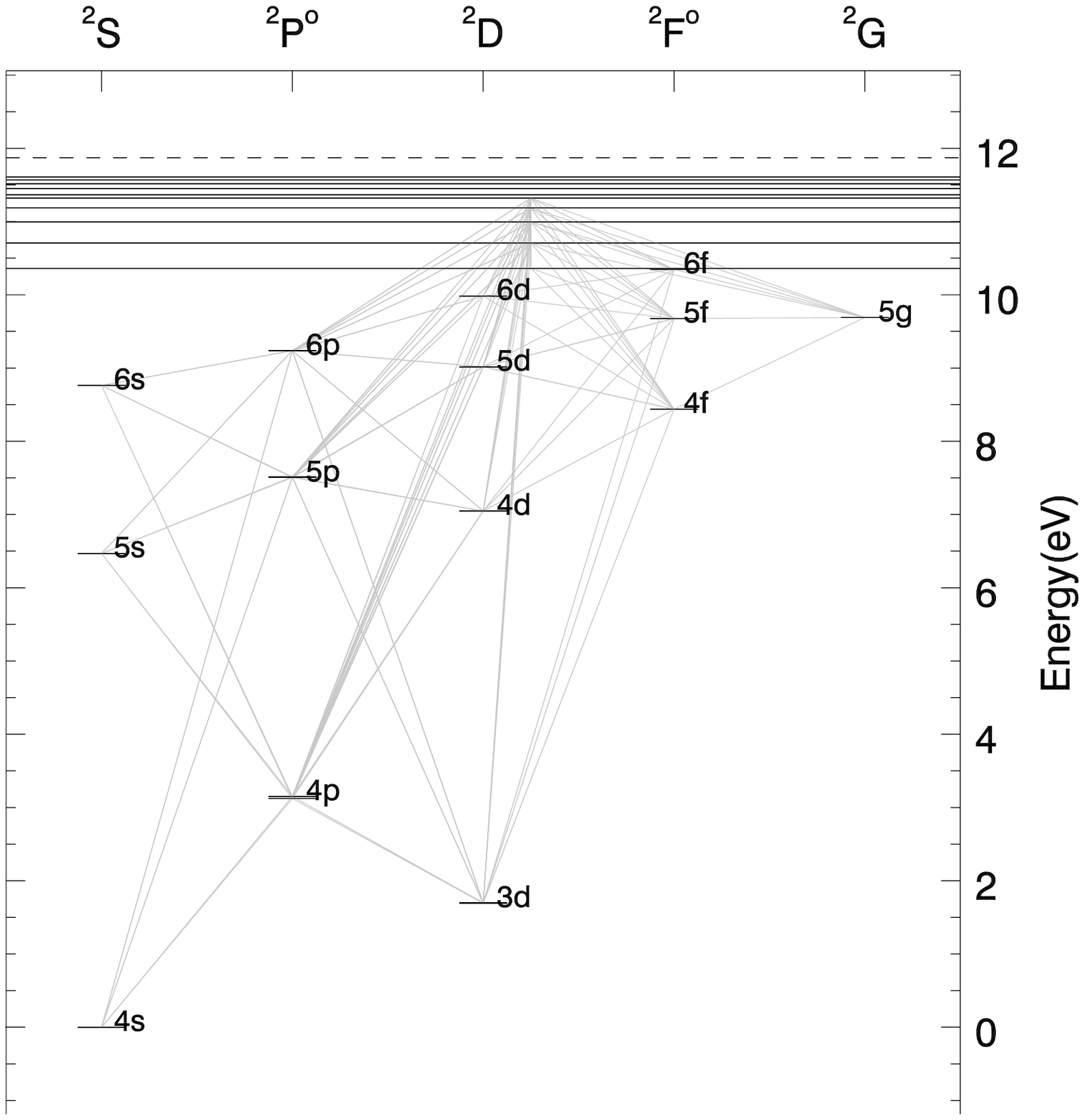}} 
};   
\node at (-0.08\textwidth,0.1\textwidth) {\scalebox{1.5}{\caii}};   
\end{tikzpicture} 
\end{tabular} 
\caption{Grotrian diagram representing the \cai\ (left) and \caii\ (right) levels and radiative transitions (grey) used in our model atom. The horizontal black solid lines represent super levels. The dashed line in each diagram shows the ionisation limit.  Only the lowest levels are labelled to ease visualisation. }\label{fig:grotrian} 
\end{figure*} 
 
 
Our model atom was built integrating radiative and collisional data mostly from the literature, and including data calculated by us. Below, we provide a description of the different radiative and collisional components adopted. 
 
\subsection{Energy levels and radiative data}

For \cai\ and \caii, energy levels were obtained from the NIST database \citep{NIST2015}. Bound-bound radiative data (f-values) were taken from the NIST and the VALD databases \citep[ from which broadening parameters were also obtained]{1995A&amp;AS..112..525P,1997BaltA...6..244R,1999A&amp;AS..138..119K,2000BaltA...9..590K}. When possible, we computed Van der Waals broadening parameters from the ABO theory \citep{1995MNRAS.276..859A,1997MNRAS.290..102B,1998PASA...15..336B,1998MNRAS.296.1057B}.

New \emph{ab initio} relativistic calculations of radiative lifetimes of bound-bound transitions of \cai\ from \cite{YU2018263} include data for 811 optically allowed transitions in LS coupling. The original information in the VALD database is mainly from \cite{K07} for \cai\ and \cite{K10} for \caii, updated with f-values of 114 bb transitions (63 for \cai\ and 51 for \caii) from \cite{DIKH}, \cite{S}, \cite{Sh}, \cite{SR}, \cite{Sm}, \cite{SN}, \cite{SG}, \cite{TB} and \cite{T}. NIST f-values are available for 134 \cai\ and 99 \caii\ transitions, compiled by \cite{1969atp..book.....W}. The preference order for the selection of f-values for transitions with multiple sources is: 
\begin{enumerate} 
\item Experimental values from VALD, 
\item \cite{YU2018263} [Only available for \cai], 
\item NIST, 
\item \cite{K07,K10}, 
\item TOPBASE. 
\end{enumerate} 
We decided on the levels to be included in the final model atom based on three criteria. First, to ensure that the gap between the highest level of \cai\ and the ground level of \caii\ was sufficiently small to guarantee that those two levels are collisionally coupled, and thus ionisation balance between \cai\ and \caii\ is not affected by any artificial energy gap (similarly for the gap between \caii\ and \caiii). Second, the fine-structure splitting of the levels is needed in order to directly compare the synthetic spectrum with the observation of lines for which fine structure produces asymmetric profiles, or that are seen as fully resolved lines. Third, the transitions used in the visual and IR to compare with observations do not come from merged levels.  
 
Our final model atom has 127 levels in total (96 for \cai, 30 for \caii\ and the ground level of \ion{Ca}{iii}). For \cai, all the triplets up to the 3d.3d($^3$F) levels are split into their fine structure components, except for the triplet 4s.4f($^3$F), which is represented by a single level. From the 4s.7s($^3$S) up to the 4s.8f ($^1$F) level, triplets are merged into a single level, except for the 4s.7p($^3$P), 4s.8d($^3$D) and the 3d.5s($^3$D) levels. The levels with $n=6, l\geq4$ are merged, as are those with $n=7, l\geq4$, and $n=8, l\geq4$. We use super levels for the levels with $9 \leq n \leq 15$. The energies for the 3d.3d($^3$P) levels (also included in the final model atom with split fine-structure components) are similar to those for the $n=12$ super level. 
 
\caii\ consists only of doublets. Fine-structure components were split for all levels up to 6p($^2$P), except for the 4f($^2$F) level. The 5f($^2$F), 5g($^2$G), 6d($^2$D), and 6f($^2$F) levels were also represented by one level each. For $n=6, l\geq4$, a single level was used; we used super levels for $7\leq n\leq15$. Isotopic splitting was considered for the \caii\ triplet lines \citep[wavelength shifts were taken from ][]{2041-8205-784-1-L17,1998EPJD....2...33N} and using the solar-system isotope abundance ratios \citep{1989GeCoA..53..197A,Asplund:2009eu}. 
 
Our final model atom has 120 bound-free transitions and 1808 bound-bound transitions (1656 of \cai\ and 152 of \caii).  When a level that has been merged is involved in a transition, the \emph{excitation} oscillator strength $f_{ij}$ is re-scaled as 
\[ 
    f_{ij}' =f_{ij}\frac{g_i}{g_i'}, 
\] 
where $f_{ij}'$ denotes the new, re-scaled, excitation oscillator strength and $g_i$, $g_i'$ are the statistical weights of the original and the new (merged) low level of the transition, respectively. Radiative data for the most relevant transitions of \ion{Ca}{} used in this work are shown in Table \ref{tab:lines}. Grotrian diagrams for \cai\ and \caii\ are shown in \fig{fig:grotrian}.

\subsection{Collisional data} 
 
The main collisional perturbers in cool stars are electrons (due to their high velocity) and hydrogen atoms (due to their abundance). We consider four collisional processes, and their inverse through detailed-balance relations: electron collisional excitation and ionisation, hydrogen collisional excitation, and charge exchange with hydrogen. In the latter, an atom of \cai\ transfers one of its electrons to the colliding hydrogen atom in its ground level and, as a result of this collision, an ionised atom (\caii) and a negative hydrogen ion (H$^-$) are produced. When levels are merged, the final transition is the sum of the transitions between the levels involved, weighted with the statistical weight of the initial levels of the transitions to merge.  Below, we provide a description of the data used to represent all these processes.  
 
 
\subsubsection{Electron collisions} 
 
Electron collisional ionisation for all levels of \caii\ and the low-lying levels of \cai\ were calculated with the hydrogenic approximation presented in \cite{2000asqu.book.....C}, based on the semi-empirical formulation in \cite{1970ARA&A...8..329B} which is original from \cite{Percival:1966tr}. For the levels above the 4s.6s($^1$S) level of \cai, we used the formula from  \cite{Vrinceanu:2005em}, suitable for electron collisional ionisation of Rydberg states of neutral atoms, and which takes into account the electronic angular momentum of the Rydberg level.  
 
The adopted electron collisional excitation rates for \cai\ are an extension of the rates calculated by \citeauthor{PhysRevA.74.052708}~(2006). This extension includes electron collisional excitation cross sections between the lowest levels of \cai\ up to 4s.8s($^1$S), using the B-spline R matrix (BSR) method. Electron collisional excitation of \caii\ for levels up to $n=8$ were taken from \cite{2007A&amp;A...469.1203M}. For higher levels, we neglected dielectronic transitions and tested both the impact parameter method (IPM, \cite{1962amp..conf..375S} for neutral species and \cite{1977MNRAS.179..275B} for positive ions) and the van Regemorter (vR) formula \citep{1970ARA&A...8..329B,1962ApJ...136..906V}, finally adopting the IPM rates. The vR formula is based on the Born approximation, which is known to overestimate electron collisional cross-sections at low energies (see \cite{1962amp..conf..375S}). Our previous work on Mg \citep{2015A&amp;A...579A..53O}  also showed that the use of the IPM (when no quantum mechanical calculations are available) reproduces the IR \ion{Mg}{I} emission lines observed in the Sun, but that if we used the vR formula instead, the above-mentioned lines cannot be reproduced.

\begin{figure}[t] 
  \vspace{-0.01\textwidth} 
    \hspace{-0.0\textwidth} \begin{tikzpicture}  
      \node[anchor=south east, inner sep=0] (image) at (0,0) { 
        \subfloat{\includegraphics[width=0.49\textwidth]{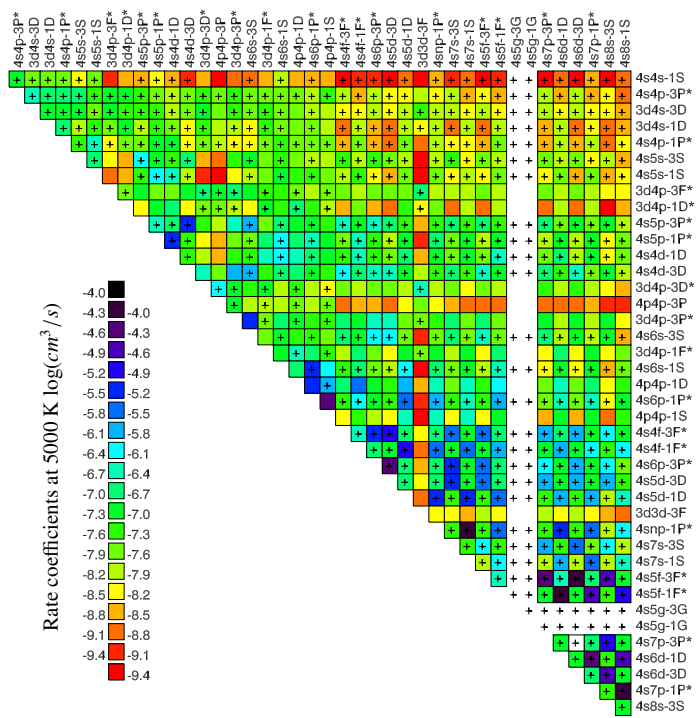}} 
      }; 
      \node[rotate=90] at (-0.46\textwidth,0.18\textwidth) {\scalebox{0.80}{Rate coefficients at 5000 K $\log(cm^3/s)$}}; 
    \end{tikzpicture}  
  \caption{Matrix with electron collisional rate coefficients at 5000~K taken from the extension of the calculations made in \cite{PhysRevA.74.052708} and used in this work. The colours represent the rate coefficients in a logarithmic scale. The crosses indicate one-electron transitions. Transitions involving the 4s.5g levels are not calculated and approximation formulas (valid only for one-electron transitions) are used.}\label{fig:zat} 
\end{figure} 
 
We decided to neglect dielectronic transitions based on comparison between one and two electron transitions calculated for this work. Figure \ref{fig:zat} shows the electron collisional rate coefficients at 5000~K obtained from quantum mechanical calculations \citep{PhysRevA.74.052708}. One-electron transitions are marked with black crosses in \fig{fig:zat}. Dielectronic transitions tend to be orders of magnitude weaker. The most clear case is with the 3d$^2$($^3$F) level, where due to its weak collisional coupling with levels with similar energies, it is the first level that shows significant departures from LTE at the deep layers of the photosphere (see Sect.  \ref{sec:computations}).

\subsubsection{Hydrogen collisions} 
 
Hydrogen collisional excitation was considered for \cai\ and neglected for \caii. Apart from \cite{2017A&A...605A..53M}, previous works on NLTE calculations for Ca that studied the effects of collisions with hydrogen made use of the Drawin formula. \cite{2017A&A...605A..53M} included detailed calculations on \cai+H collisions by \cite{2016A&A...587A.114B} and the Drawin formula for \caii+H collisions in their model atom. The Drawin formula is used to treat excitation of atoms due to collisions with hydrogen only for optically allowed transitions; although it has been shown that it overestimates collision rates by orders of magnitude \citep{2011A&A...530A..94B} and often an arbitrary scale-factor is used. We adopted the data from \cite{2016PhRvA..93d2705B, 2017PhRvA..95f9906B}, who uses a new theoretical method based on an asymptotic two-electron model of ionic-covalent interactions for the calculation of \mbox{hydrogen-atom} collisional excitation and charge-transfer processes. We included collisional excitation  between the lowest levels of \cai\, up to 4s.6s($^3$S), but ignoring two-electron (de)excitation processes.  Charge-transfer processes with H involving the same levels of \cai\ as above and the three lowest levels of \caii\, were considered. Again, only transitions involving the removal of one of the valence electrons were considered. Hydrogen collisional excitations between levels above 4s.6s ($^3$S) were treated using the formula for Rydberg-neutral inelastic collisions from \cite{Kaulakys:1986tl}, disregarding dielectronic transitions.

\begin{figure*}[ht] 
  \centering 
    \hspace{-0.05\textwidth} \begin{tikzpicture}  
      \node[anchor=south east, inner sep=0] (image) at (0,0) { 
        \subfloat{\includegraphics[width=1.0\textwidth]{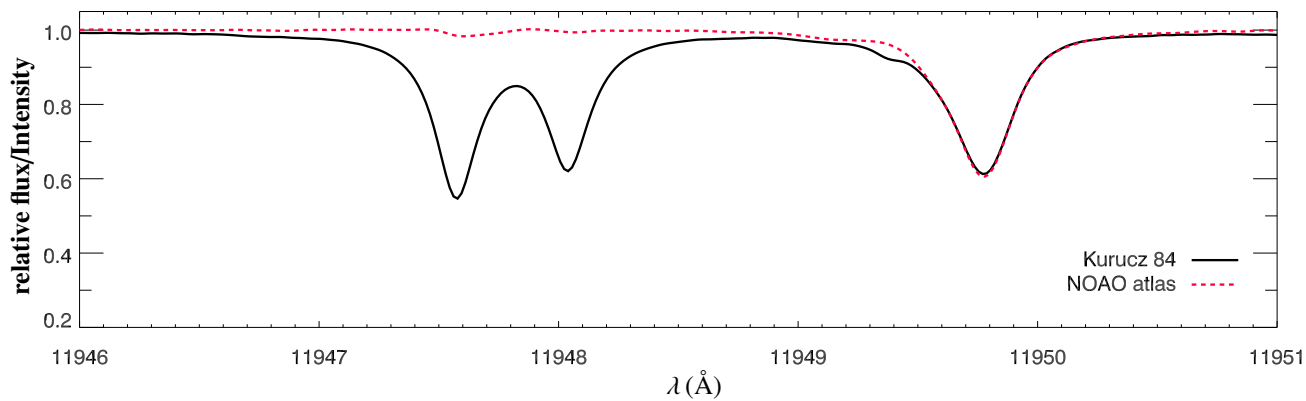}} 
      }; 
      \node[rotate=90] at (-0.93\textwidth,0.18\textwidth) {\scalebox{0.99}{\bf relative flux/Intensity}}; 
      \node at (-0.45\textwidth,0.03\textwidth) {\scalebox{0.99}{\bf $\lambda\,(\AA)$}} ;     
    \end{tikzpicture}  
  \caption{Comparison of the observations in part of the overlapping region between the two atlases used for the Sun: \cite{kursun84} for the visual region (solid black lines) and \cite{sun_ir} for the IR (red dashed lines). The later atlas has telluric lines removed and measures the disc centre intensity of the Sun. The wavelengths were shifted in order to match the observations.}\label{fig:compare_vir} 
\end{figure*}

\section{Reference stars}

In order to test our model atom for Ca we used three benchmark stars with observations of exceptional quality and well-known parameters: The Sun, Arcturus, and Procyon. For the Sun, we adopted \teff=5772 K, \logg[cm/s$^2$]=4.44, and \feh=0.0~dex, with the solar element abundances from \cite{2005ASPC..336...25A} and micro-turbulent velocity V$_{mic}$=1.1 km\,s$^{-1}$. For Arcturus, we adopted \teff/\logg/\feh~=~4300/1.5/-0.5 and  V$_{mic}$=1.6 \citep{2011ApJ...743..135R}. In the case of Procyon the parameters were \teff/\logg/\feh~=~6530/4.00/0.0 and V$_{mic}$=2.0 \citep{2002ApJ...567..544A}. These parameters, together with the derived calcium abundance, \Aca{}{}, are summarised in Table \ref{tab:observations}.

Calcium lines are detectable in a wide wavelength range of the spectrum of late-type stars. The variety in strength and excitation level makes calcium a good candidate for testing various NLTE processes. The UV continuum forms in higher atmospheric layers than the visual and near-infrared (NIR) continuum, thus lines in this region are typically affected by severe departures from LTE. Strong lines form their wings and cores at very different atmospheric depths, and therefore different NLTE effects can be at play shaping them. 
 
For the Sun, we adopted the atlas from \cite{kursun84}; it covers an area from 3\,000 to 13\,000~\AA\ and has a resolution of R$\approx$400\,000. We compared this atlas with the PEPSI solar atlas \citep{2018A&A...612A..44S}, which has lower resolution and therefore shallower line cores (see Fig. 4). Good agreement is however found after smoothing the higher-resolution atlas to match the PEPSI resolution.

For the NIR region we used the atlas from \cite{sun_ir} that spans between 11\,000 and 50\,000~\AA\ with a resolving power $\approx$300\,000 and measures the disc-centre intensities of the Sun with telluric lines removed. In the overlap between both atlases there are two clean \caii\ lines: 11\,838 and 11\,949~\AA. Figure \ref{fig:compare_vir} shows part of the overlapping region around the 11\,949 \caii\ line.

For Arcturus, we adopted the high-resolution atlases in the visible \citep[R$\approx$150\,000]{2000vnia.book.....H}, spanning between 3\,727 and 9\,300~\AA,  and in the IR \citep[R$\approx$100\,000]{1995PASP..107.1042H}, spanning between 9\,129 and 53\,000~\AA.  
 
\begin{figure}[t] 
  \begin{tabular}{c} 
    \hspace{-0.1\textwidth} \begin{tikzpicture} 
      \node[anchor=south east, inner sep=0] (image) at (0,0) { 
        \subfloat{\includegraphics[width=0.58\textwidth]{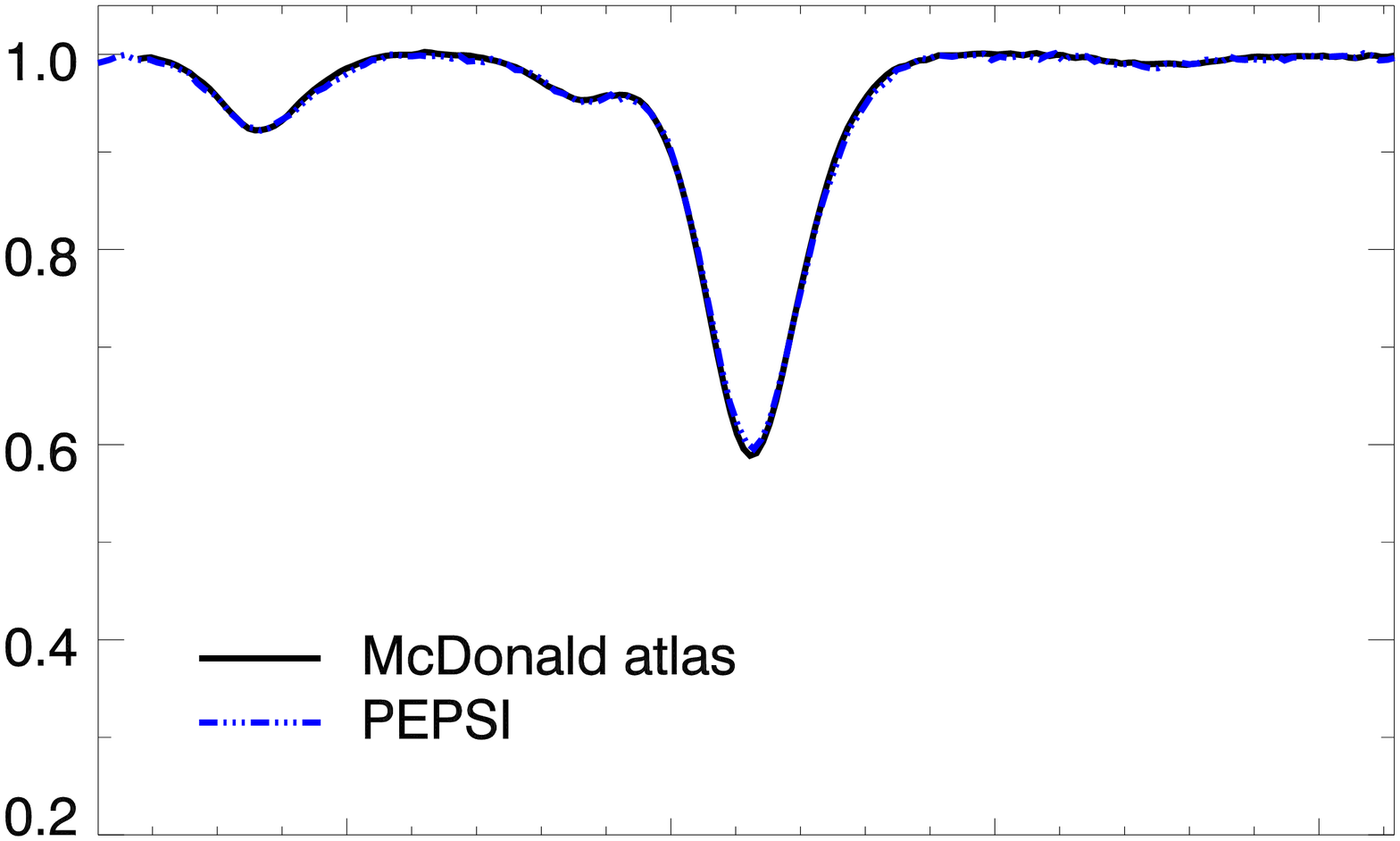}} 
      }; 
      \node at (-0.1\textwidth,0.1\textwidth) {{Procyon}} ; 
    \end{tikzpicture} \\[-0.14\textwidth] 
    \hspace{-0.1\textwidth} \begin{tikzpicture} 
      \node[anchor=south east, inner sep=0] (image) at (0,0) { 
        \subfloat{\includegraphics[width=0.58\textwidth]{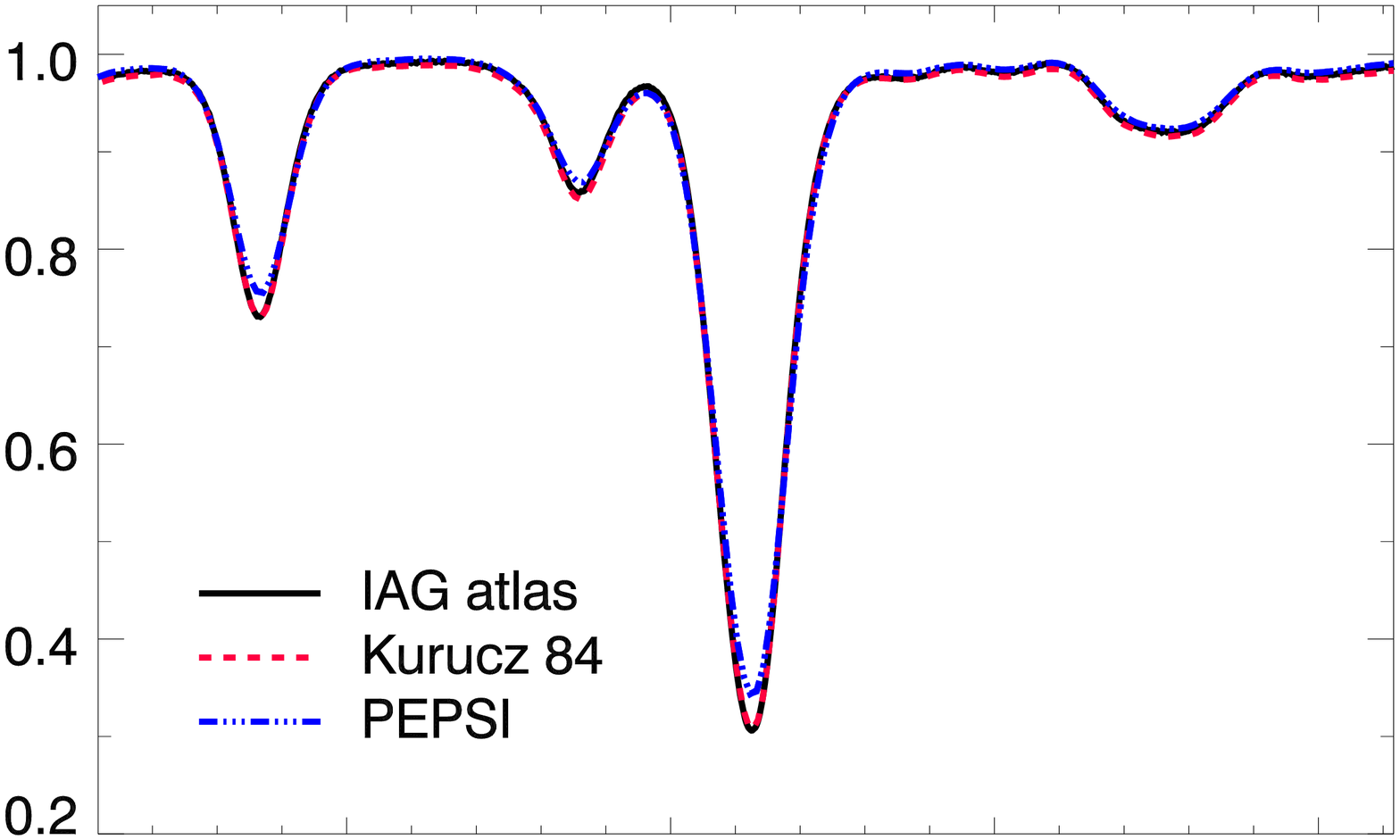}} 
      }; 
      \node at (-0.1\textwidth,0.1\textwidth) {{Sun}} ; 
      \node[rotate=90] at (-0.51\textwidth,0.22\textwidth) {\scalebox{0.90}{relative flux}}; 
    \end{tikzpicture} \\[-0.14\textwidth] 
    \hspace{-0.1\textwidth} \begin{tikzpicture} 
      \node[anchor=south east, inner sep=0] (image) at (0,0) { 
        \subfloat{\includegraphics[width=0.58\textwidth]{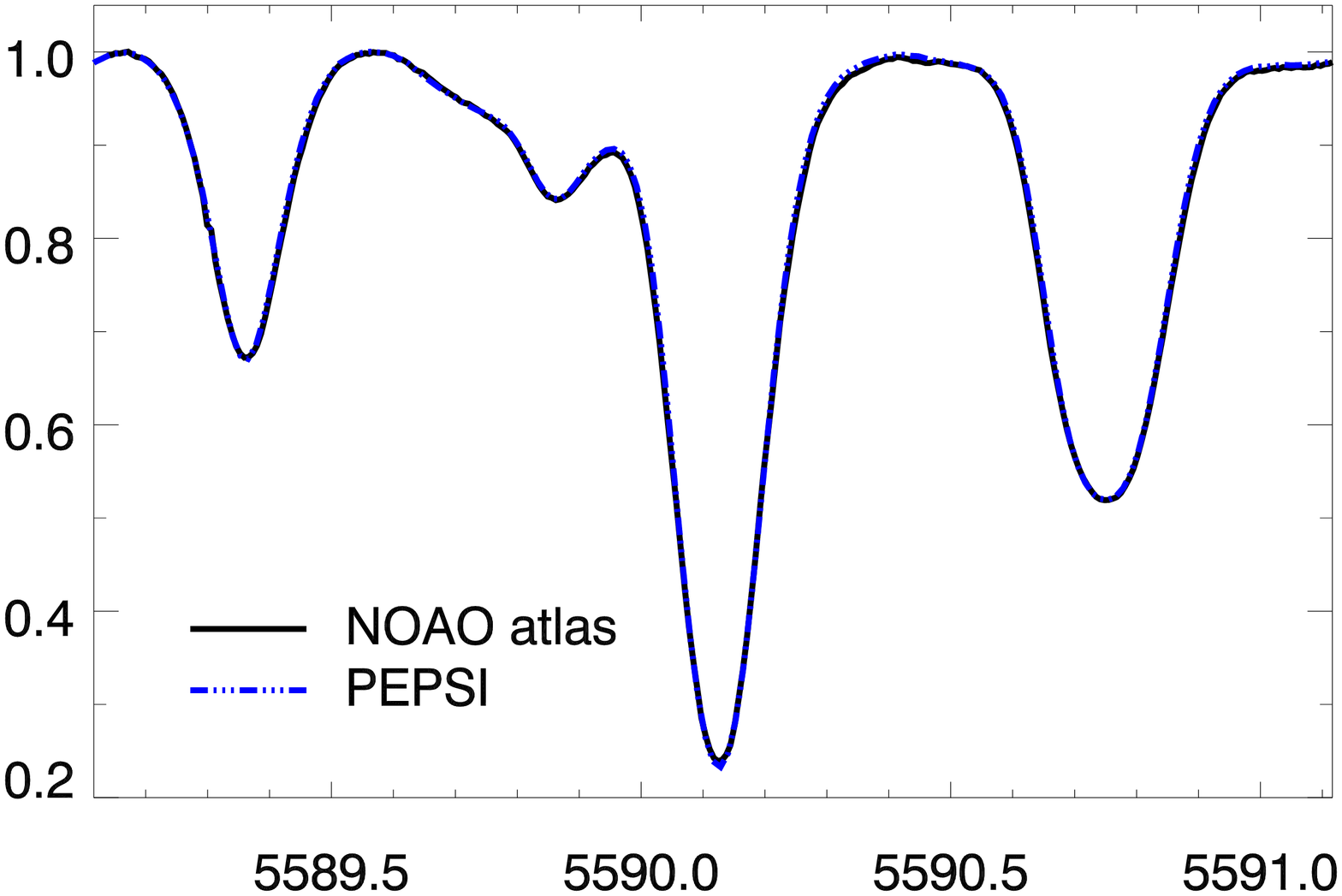}} 
      }; 
      \node at (-0.1\textwidth,0.1\textwidth) {{Arcturus}} ; 
      \node at (-0.25\textwidth,0.03\textwidth) {\scalebox{0.90}{$\lambda\,(\AA)$}} ;     
    \end{tikzpicture}\\[-0.03\textwidth] 
  \end{tabular} 
  \caption{Comparison of the observations of the \cai\ 5590~\AA\ line between popular atlases of the three studied stars. PEPSI atlases are from  \cite{2018A&A...612A..44S} and \cite{2018A&A...612A..45S}. IAG atlas comes from \cite{2016A&A...587A..65R}, NOAO atlas for Arcturus comes from \cite{2000vnia.book.....H} and McDonald data come from \cite{2002ApJ...567..544A}. The atlases were wavelength shifted in order to match the air wavelength of the 5590~\AA\ \cai\ line.}\label{fig:compare_atlas} 
\end{figure}

The Procyon observations are from PEPSI \citep{2018A&A...612A..45S}. The PEPSI spectrum covers wavelengths from 3\,800 to 9\,100~\AA\ at R$\approx$220\,000, therefore including the \caii\ triplet. We compared them with observations from \cite{2002ApJ...567..544A}, which span from 4\,560 to 7\,400~\AA, and found excellent agreement, as illustrated in \fig{fig:compare_atlas}.

\section{Computations}\label{sec:computations}

The calculations of LTE/NLTE synthetic line profiles and populations were performed using the code {\tt MULTI} (version 2.3) \citep{1986UppOR..33.....C,1992ASPC...26..499C}, which adopts the plane parallel approximation on its radiative transfer and statistical equilibrium calculations. The same is true for the 1D LTE model atmospheres we employ in our calculations (see Sect.  \ref{atmospheres}).  
 
{\tt MULTI} uses the trace-element approach, where it is assumed that the departures from LTE of the level populations of the element under study do not affect the atmospheric structure or the background opacities. Therefore, atmospheric parameters and background opacities are held fixed. The latter assumption may be justified since the calcium contribution to the background opacity and the free-electron reservoir is < 10\% of the contribution made by more-abundant elements such as Fe, Si, or Mg. We adopted the same background line opacity data used in the MARCS grid presented in \cite{Gustafsson:2008df}, re-sampled on 10\,300 frequency points. 
 
The radial-tangential formulation \citep{1975ApJ...202..148G} was implemented to represent macro-turbulence and we allowed it to vary at the same time as the calcium abundance in order to fit the observations. The values adopted for vsini were extracted from the literature: 5.4, 1.6, and 4.0 km\,s$^{-1}$ for Procyon \citep{2010A&A...520A..79M}, the Sun \citep{2012MNRAS.422..542P}, and Arcturus \citep{2008AJ....135..209M}, respectively. 
 

\subsection{Stellar atmospheric models} 
\label{atmospheres} 

\begin{figure}[t] 
\hspace{-0.05\textwidth}\begin{tikzpicture} 
\node[anchor=south east, inner sep=0] (image) at (0,0) {  
  \subfloat{\includegraphics[width=0.55\textwidth]{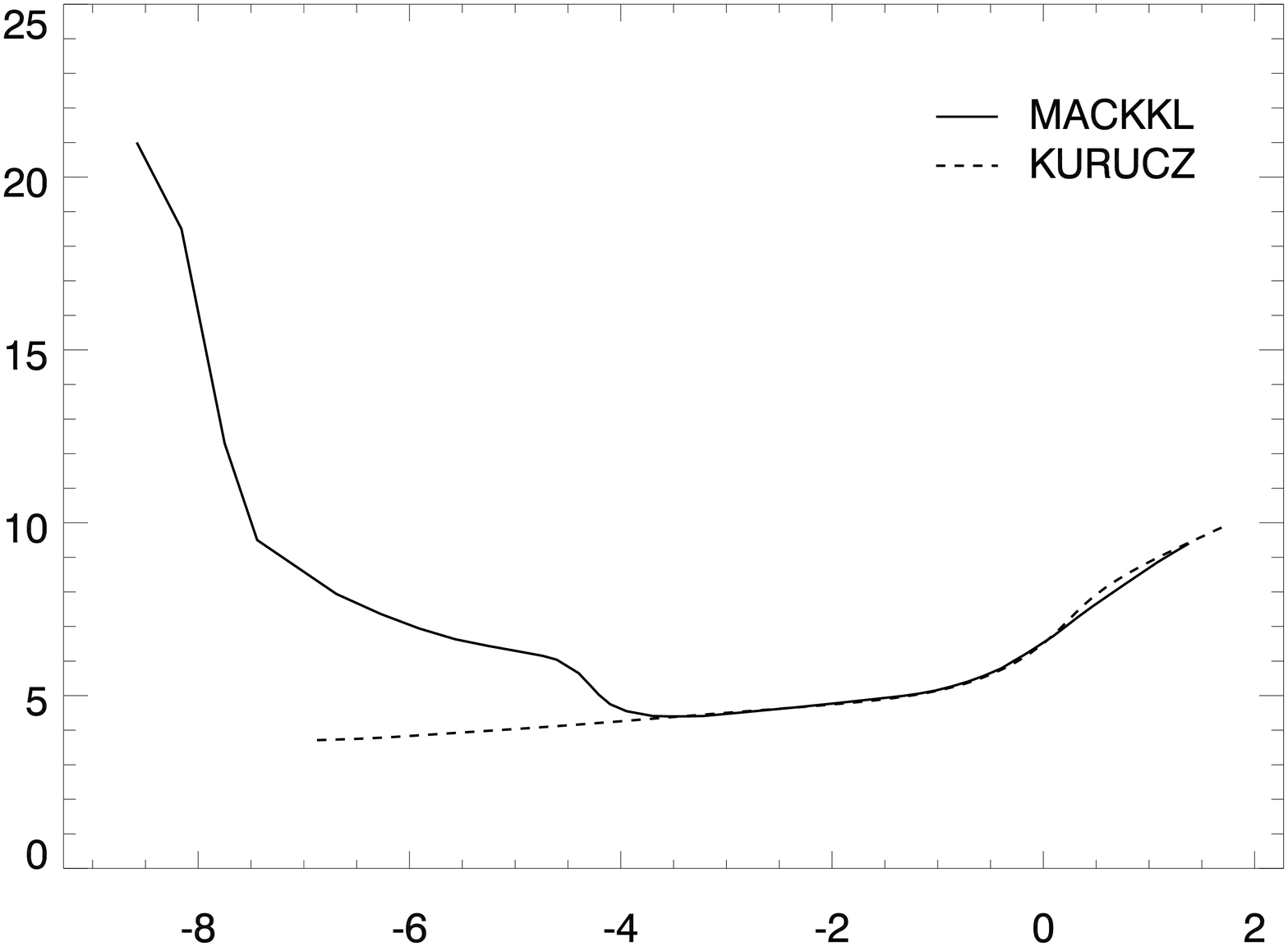}} 
}; 
\node[rotate=90] at (-0.478\textwidth,0.22\textwidth) {\scalebox{0.9}{\bf Temperature ($1\,000$ K)}}; 
\node at (-0.25\textwidth,0.03\textwidth) {\scalebox{1.0}{\bf \taufive}} ; 
\end{tikzpicture}\\[-0.05\textwidth] 
\caption{Temperature structure of the two solar models used in this work. The MACKKL model includes the lower chromosphere.}\label{fig:sun_atmos} 
\end{figure}

In this work we used Kurucz model atmospheres computed with ATLAS9 \citep{1993KurCD..13.....K}, using exactly the same setup described by \cite{2012AJ....144..120M}. For the centre-to-limb variation (CLV) analysis, we also tested the \emph{photospheric reference model} from \cite{1986ApJ...306..284M}, as this is a semi-empirical model intended to represent the quiet Sun and the average thermodynamical properties of the chromosphere. We refer to this model as the MACKKL model, and its thermal structure is shown and compared with its Kurucz counterpart in \fig{fig:sun_atmos}. 
 
 {\tt MULTI} accepts input depth points defined in column mass or optical depth at 5000 \AA\ (\taufive) in the model atmosphere, but the calculations are performed in a \taufive\ scale. If the input is provided in a column mass scale, the \taufive\ data points are calculated by integration of the density multiplied by the absorption coefficient using the trapezoidal rule. We experimented using both \taufive\ and column-mass formats in the input Kurucz model atmosphere for {\tt MULTI}, and found that adopting a column mass scale for the input leads to optical depths compatible with the background opacities used in the calculations. We also found that using column mass on the input model atmosphere results in structures on an optical depth scale that are in good agreement with the MARCS models \citep{Gustafsson:2008df}, which were calculated with the same background opacities used in this work. 
 
Given that the main NLTE effect visible in the stellar spectra in this study is the deepening of the core of Ca lines, and the formation of these cores occurs in high atmospheric layers, we decided to adopt Kurucz model atmospheres over MARCS due to their larger extent. For example, for the solar models available for the Sun, the ranges in optical depth are $\log_{10}$(\taufive)=[-4.9,1.8] and [-6.9,1.8] for the MARCS and Kurucz model atmospheres, respectively. 
 
In LTE the cores of the \caii\ IR-triplet lines form at the top boundary of the MARCS model atmospheres, while for the Kurucz models the same region is well under the outer limit. The NLTE calculations make this situation worse, shifting the formation of the core of these lines to $\log_{10}$(\taufive) $\sim-6$ in the Kurucz model.

\subsection{NLTE populations and spectral synthesis}

\begin{figure}[t] 
\begin{tabular}{c} 
\begin{tikzpicture} 
\node[anchor=south east, inner sep=0] (image) at (0,0) {  
  \subfloat{\includegraphics[width=0.45\textwidth]{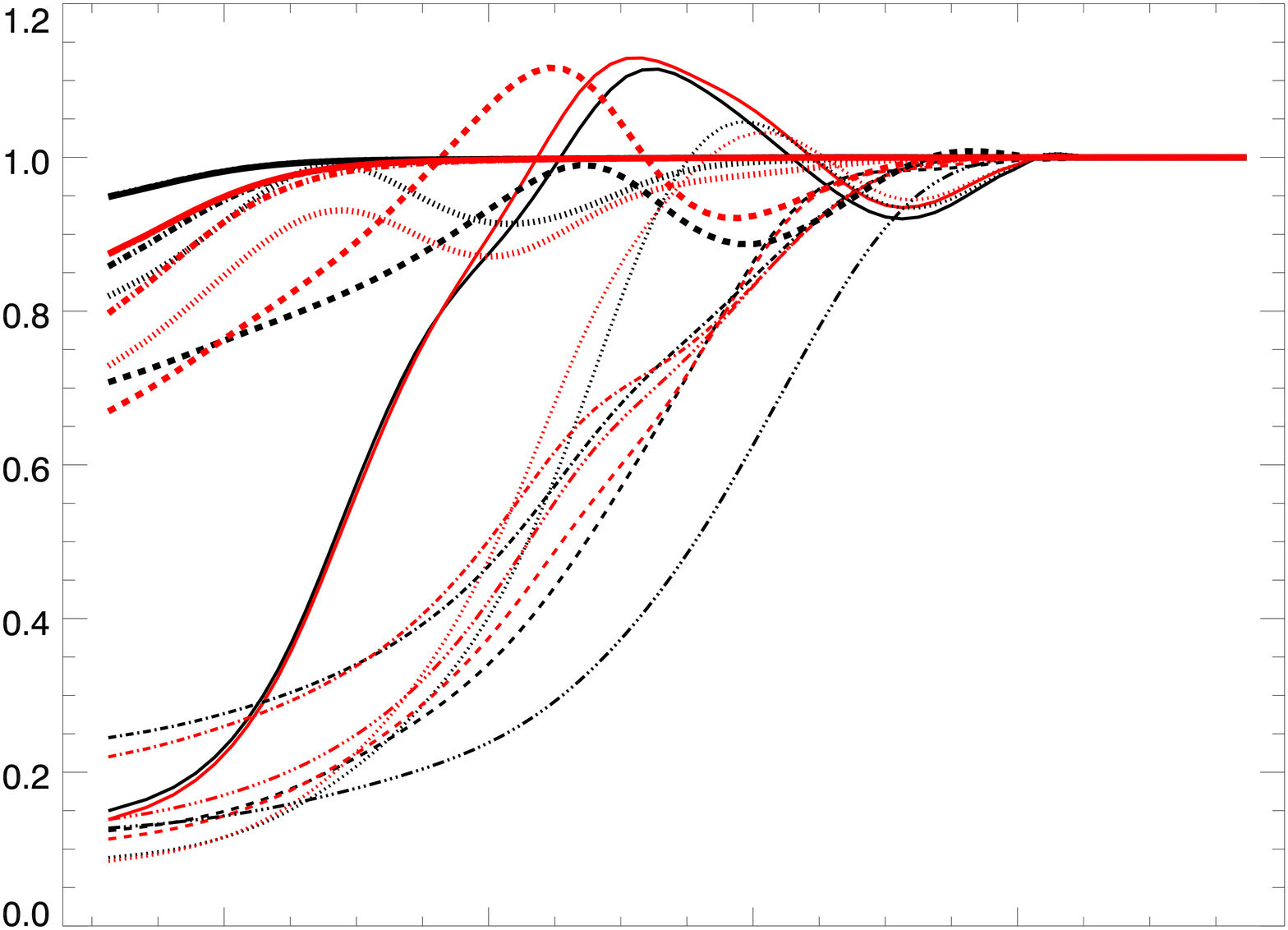}} 
}; 
\node[anchor=east] at (-0.01\textwidth,0.30\textwidth) {\scalebox{1.1}{Procyon}}; 
\node at (-0.46\textwidth,0.18\textwidth) {\scalebox{0.90}{$b$}}; 
\node[anchor=west] at (-0.15\textwidth,0.10\textwidth) {\scalebox{0.90}{\color{red}{e col: vR+A}}}; 
\node[anchor=west] at (-0.15\textwidth,0.07\textwidth) {\scalebox{0.90}{{e col: updated data}}}; 
\end{tikzpicture} \\[-0.01\textwidth] 
\begin{tikzpicture} 
\node[anchor=south east, inner sep=0] (image) at (0,0) {  
  \subfloat{\includegraphics[width=0.45\textwidth]{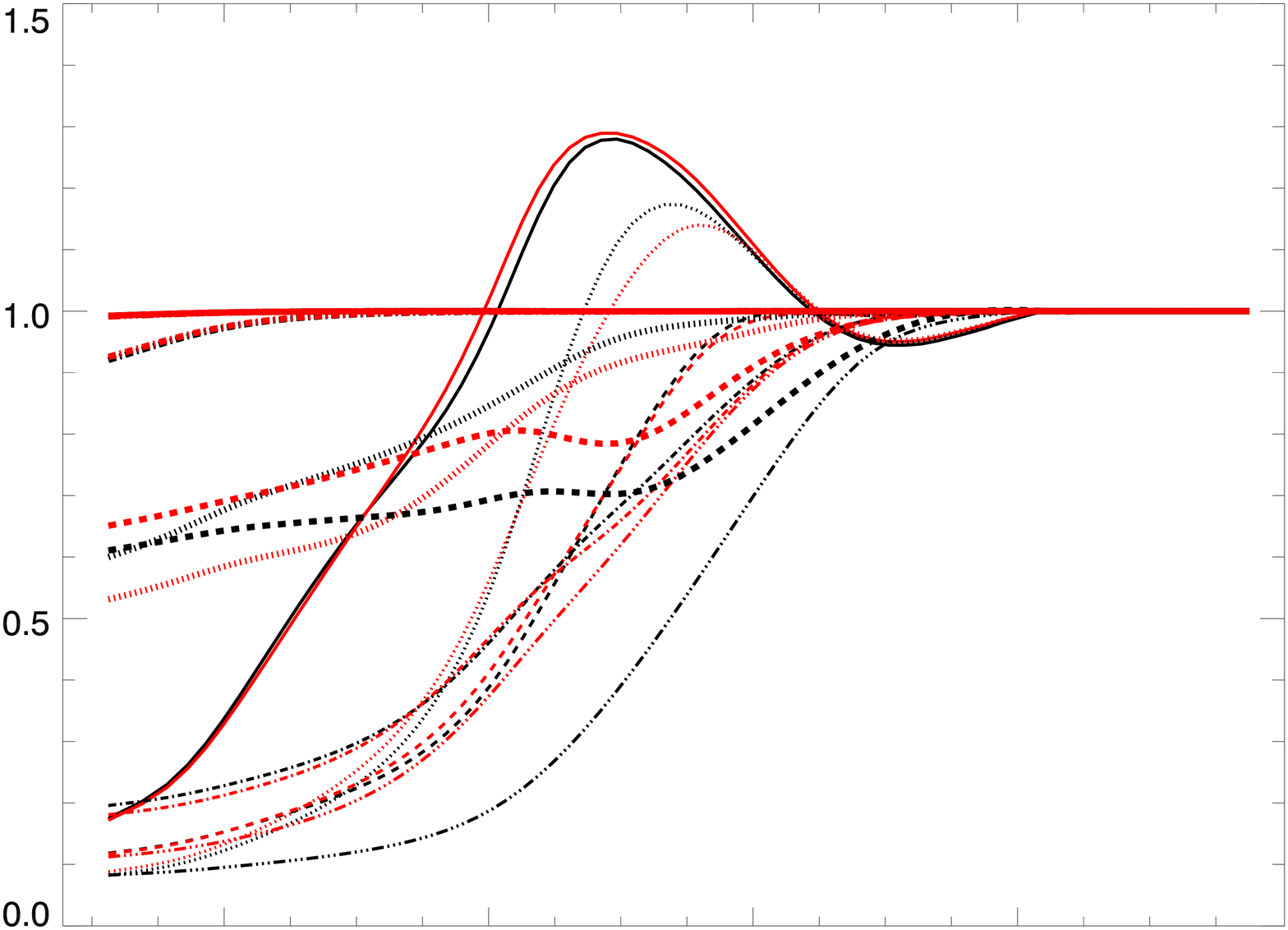}} 
}; 
\node[anchor=east] at (-0.01\textwidth,0.30\textwidth) {\scalebox{1.1}{Sun}}; 
\node at (-0.46\textwidth,0.16\textwidth) {\scalebox{0.90}{$b$}}; 
\end{tikzpicture} \\[-0.01\textwidth]  
\begin{tikzpicture} 
\node[anchor=south east, inner sep=0] (image) at (0,0) {  
  \subfloat{\includegraphics[width=0.45\textwidth]{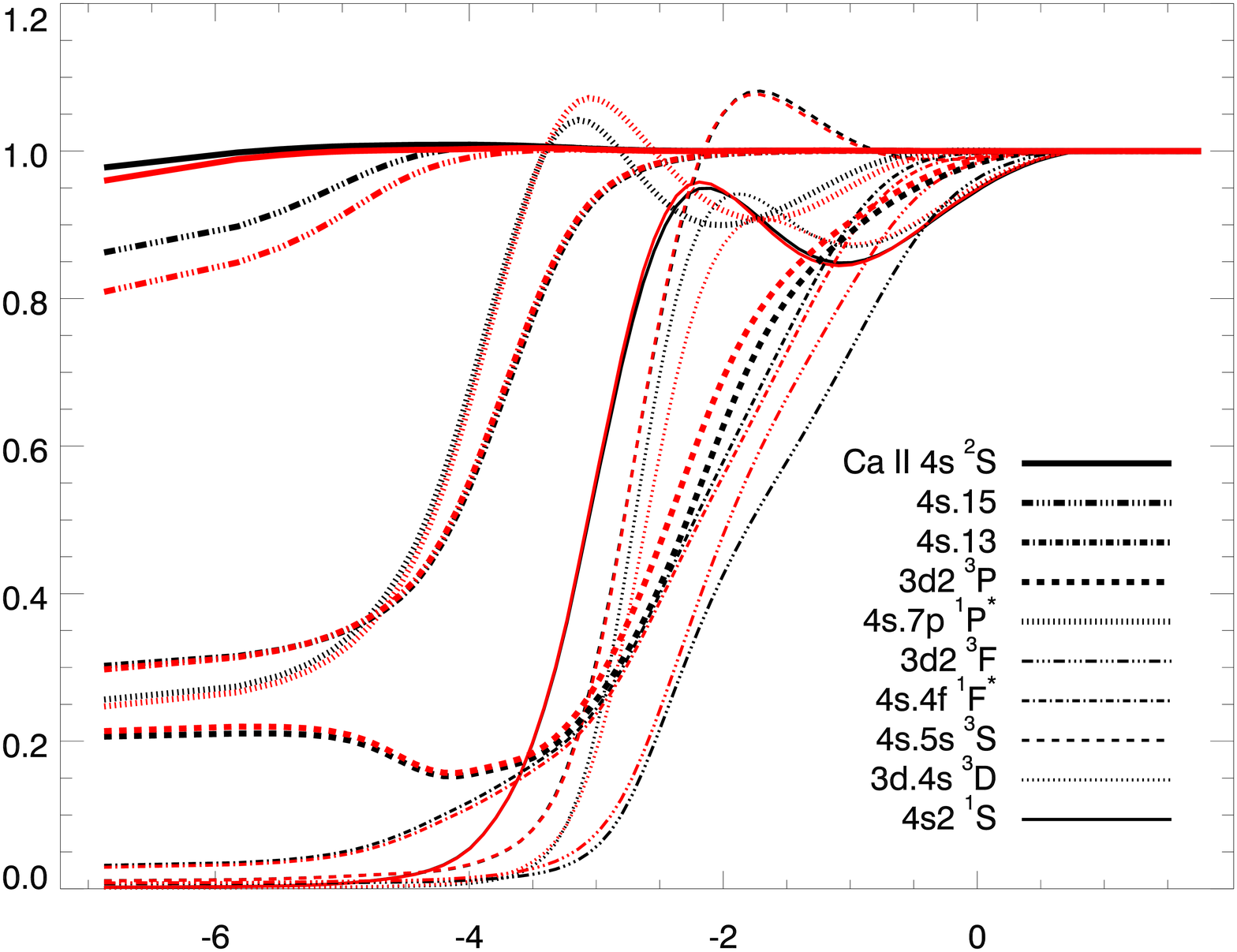}} 
}; 
\node[anchor=east] at (-0.01\textwidth,0.32\textwidth) {\scalebox{1.1}{Arcturus}}; 
\node at (-0.46\textwidth,0.17\textwidth) {\scalebox{0.90}{$b$}}; 
\node at (-0.235\textwidth, -0.01\textwidth) {\scalebox{0.90}{$\log\tau_{5000}$}} ; 
\end{tikzpicture} \\[-0.01\textwidth] 
\end{tabular} 
\caption{Departure coefficients ($b=n_{\rm{NLTE}}/n_{\rm{LTE}}$) of selected levels of \cai\ and the ground level of \caii\ using different electron collisional data. Line symbols represent different levels; In red are the departure coefficients obtained when the vR+A data are used for electron collisions (see text), and in black are the $b$ coefficients obtained when updated electron collisional data are used.}\label{fig:bcoef} 
\end{figure}

In order to test the impact of the electron collisional data adopted in this work, we compared our results with those obtained with standard formulae, namely those by van Regemorter (vR) and Allen \cite[taken from][]{2000asqu.book.....C} for electron collisional excitation and ionisation, respectively. This set of electron collisional data will be referred to as vR+A; in contrast, our default atom contains electron collisional data from quantum mechanical calculations for excitation and a more sophisticated formula \citep{Vrinceanu:2005em} for ionisation.

The use of the vR+A data tends to increase the population of the low-lying levels of \cai\ at line-formation layers (which for Procyon and the Sun implies larger departures from LTE) while intermediate and high levels of \cai\ tend to decrease their population when compared with NLTE calculations using updated electron collisional data. A notable case is the 3d$^2$($^3$P) levels where, as mentioned earlier, quantum mechanical calculations of electron collisional cross sections show a weak collisional coupling between this level and other levels with similar energy while the use of vR+A rates couples the 4s.4f($^1F^*$) and 3d$^2$($^3$P) levels leaving them in relative LTE. The thick three-dot-dash lines in \fig{fig:bcoef} represent the departure coefficients\footnote{Departure coefficient of level $i$ is \[b_i=N_i/N_i^*,\] where $N_i$ is the NLTE population and $N_i^*$ the LTE population of level~$i$.} of the n=15 super-level ($b_{n=15}$). For the Sun and Procyon, these levels couple with the ground level of \caii\ (which is thermalised). In the case of Arcturus, the $n=13$ level decouples from the ground level of \caii\ \mbox{at $\log\,\tau_{5000}\lesssim-2.5$} while the $n=15$ level (the highest level adopted for \cai\ in our model atom) remains coupled with the ground level of \caii\ in most of the atmosphere, including the line-formation region, up to $\log\tau_{5000}\sim-4$. In reality, the highest levels of \cai\ and the continuum (i.e. the ground level of \caii) are collisionally coupled and this must be reflected in the modelling of the populations. As mentioned earlier, this is one of the criteria used to select the levels of our final model atom.

\section{Comparisons with observations}

We compare high-quality observations of the Sun, Arcturus, and Procyon with synthetic spectra of \cai\ and \caii\ lines in LTE and NLTE conditions. Due to the large number of Ca lines present in the spectra of late-type stars we can take advantage of the wide spectral range offered by the available atlases to investigate NLTE effects. Additionally, we examine the CLV in the solar spectrum, which is also a useful diagnostic for departures from LTE.  

\subsection{Reference stars}
 
\begin{figure*} 
  \centering 
   \begin{tabular}{c c c }\\[-0.1\textwidth] 
    \hspace{-0.07\textwidth}\begin{tikzpicture} 
      \node[anchor=south west, inner sep=0,rotate=-90] (image) at (0,0) { 
        \subfloat{\includegraphics[width=1.0\textwidth]{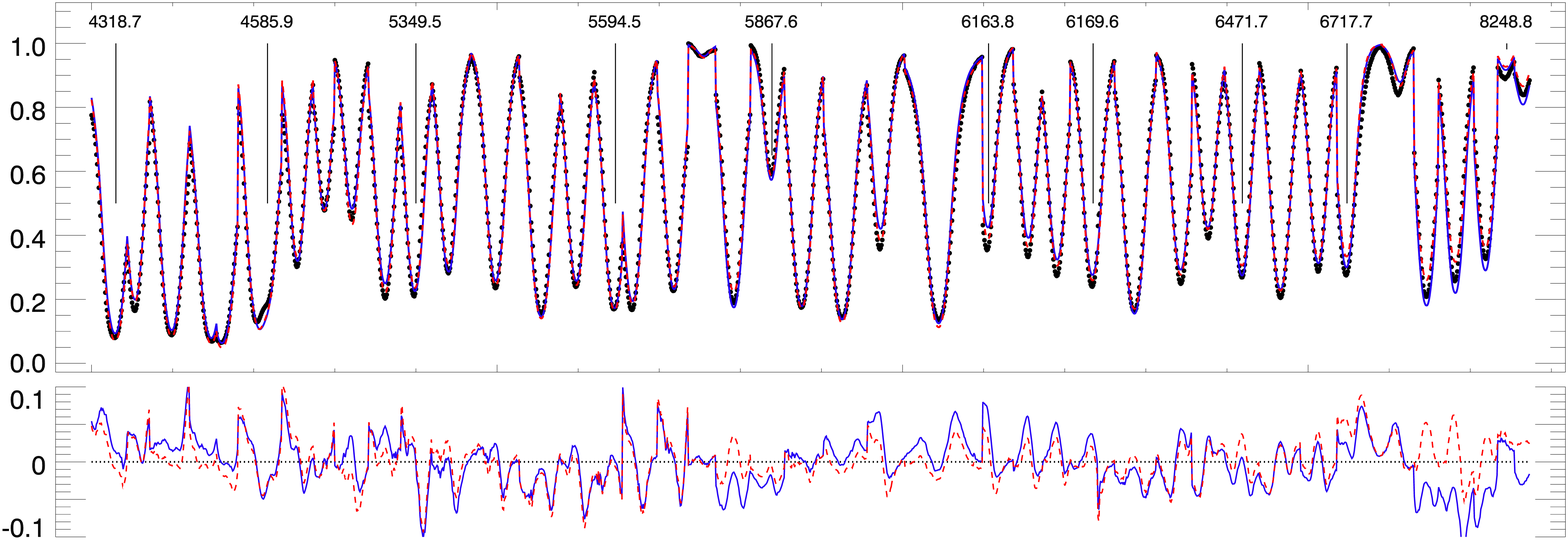}} 
      }; 
    \end{tikzpicture} 
    & 
    \hspace{-0.13\textwidth}\begin{tikzpicture} 
      \node[anchor=south west, inner sep=0,rotate=-90] (image) at (0,0) { 
        \subfloat{\includegraphics[width=1.0\textwidth]{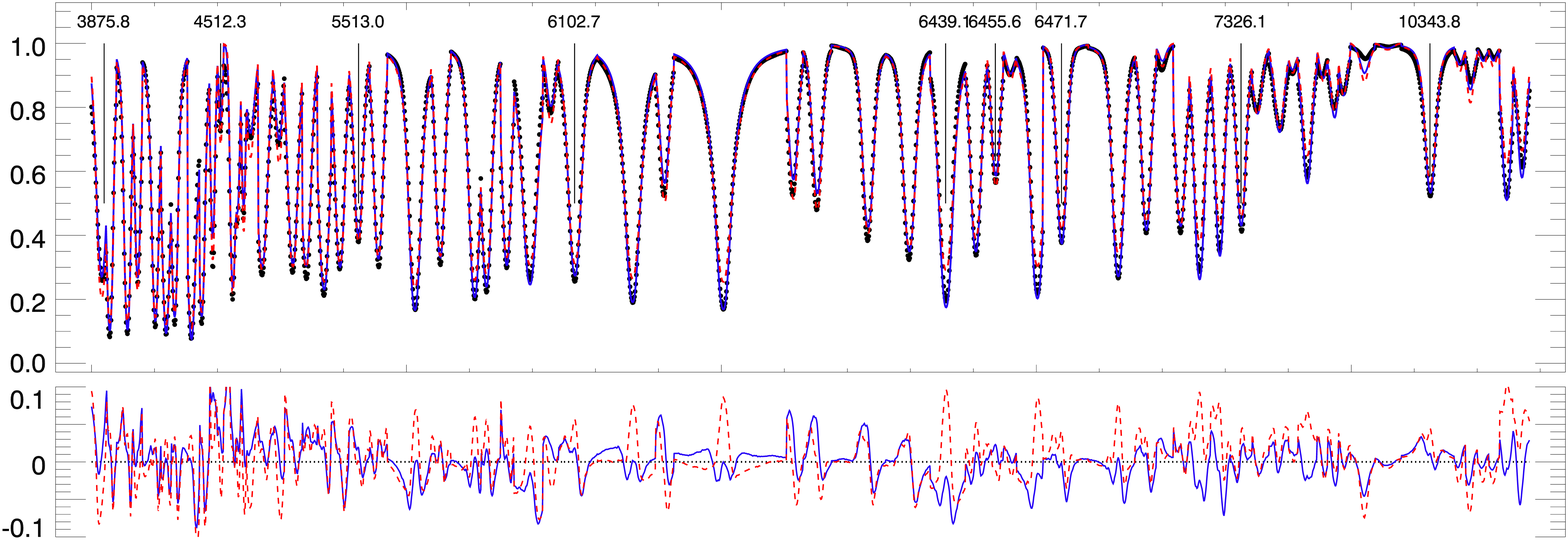}} 
      }; 
    \end{tikzpicture} 
    & 
    \hspace{-0.13\textwidth}\begin{tikzpicture} 
      \node[anchor=south west, inner sep=0,rotate=-90] (image) at (0,0) { 
        \subfloat{\includegraphics[width=1.0\textwidth]{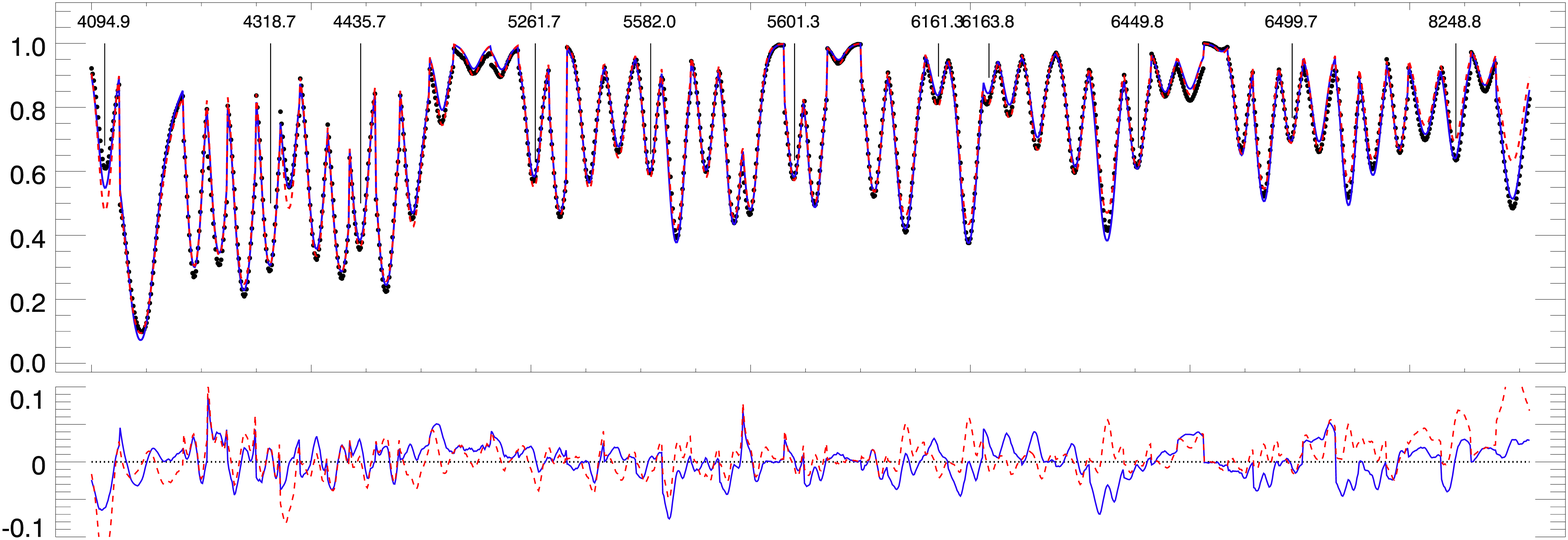}} 
      }; 
      \node[rotate=-90] at (0.23\textwidth,-0.92\textwidth) {\scalebox{0.8}{ Obs }}; 
      \draw[black,fill=black] (0.23\textwidth,-0.88\textwidth) circle (0.003\textwidth);  
     \node[rotate=-90] at (0.21\textwidth,-0.92\textwidth) (nl) {\scalebox{0.8}{\color{blue} NLTE}};       
     \draw [line width=0.4mm,blue] (nl) -- ++(90:0.07\textwidth) ;  
     \node[rotate=-90] at (0.19\textwidth,-0.92\textwidth) (lt) {\scalebox{0.8}{\color{red}  LTE}};       
     \draw [line width=0.5mm,red,dashed] (lt) -- ++(90:0.07\textwidth) ; 
    \end{tikzpicture} 
    \\[-0.08\textwidth] 
    \hspace{-0.05\textwidth}\begin{tikzpicture} 
     \node[anchor=south west, inner sep=0,rotate=-90] (image) at (0,0) { 
      \subfloat{\includegraphics[width=0.45\textwidth]{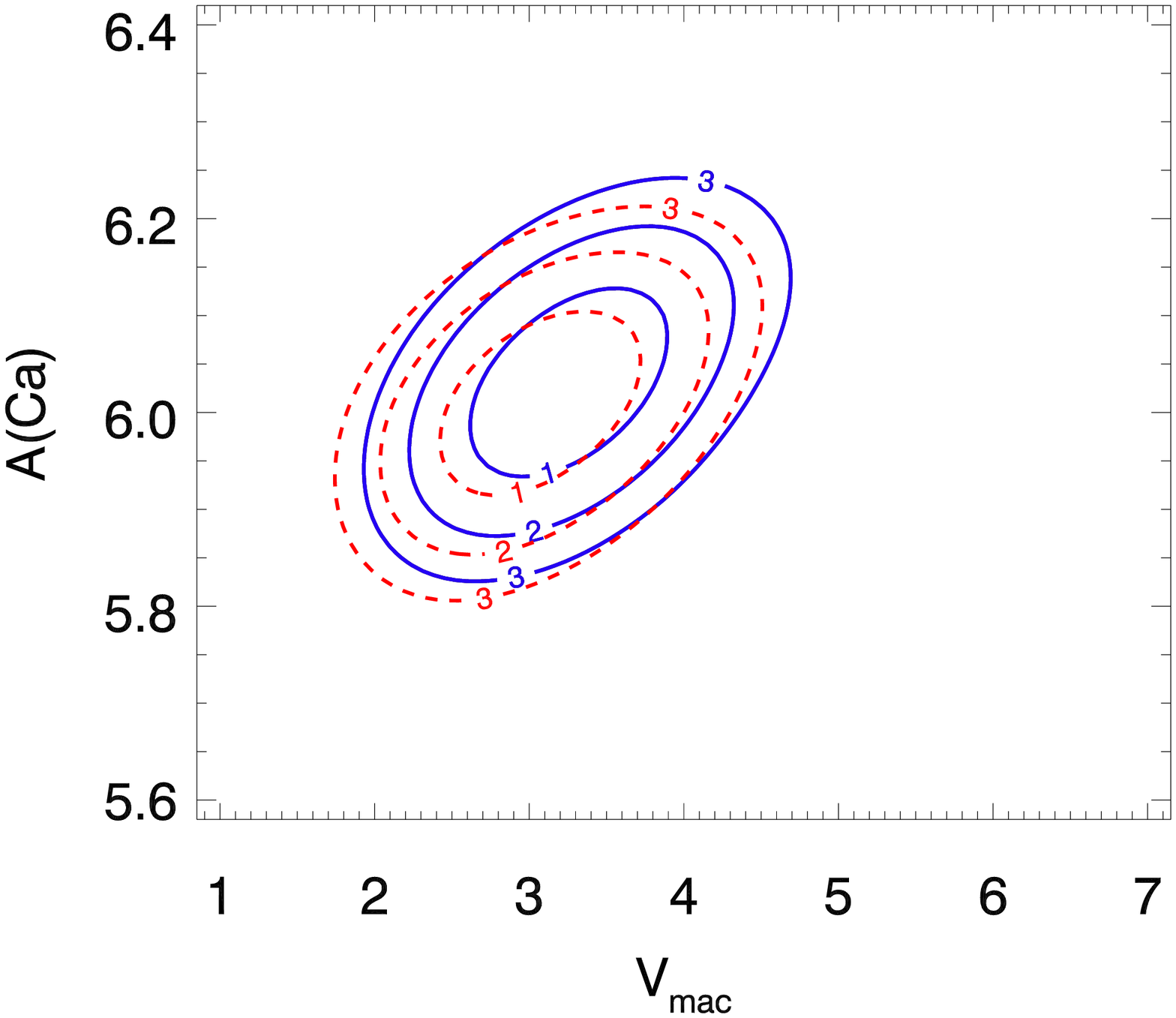}} 
     }; 
      \node[rotate=-90] at (0.30\textwidth,-0.16\textwidth) {Arcturus}; 
     \node[rotate=-90] at (0.11\textwidth,-0.38\textwidth) (nt) {\scalebox{0.8}{\color{blue} NLTE}}; 
     \draw [line width=0.4mm,blue] (nt) -- ++(90:0.05\textwidth) ; 
     \node[rotate=-90] at (0.09\textwidth,-0.38\textwidth) (lt) {\scalebox{0.8}{\color{red}  LTE}}; 
     \draw [line width=0.4mm,red,dashed] (lt) -- ++(90:0.05\textwidth);  
    \end{tikzpicture}  
    & 
    \hspace{-0.1\textwidth}\begin{tikzpicture} 
     \node[anchor=south west, inner sep=0,rotate=-90] (image) at (0,0) { 
      \subfloat{\includegraphics[width=0.45\textwidth]{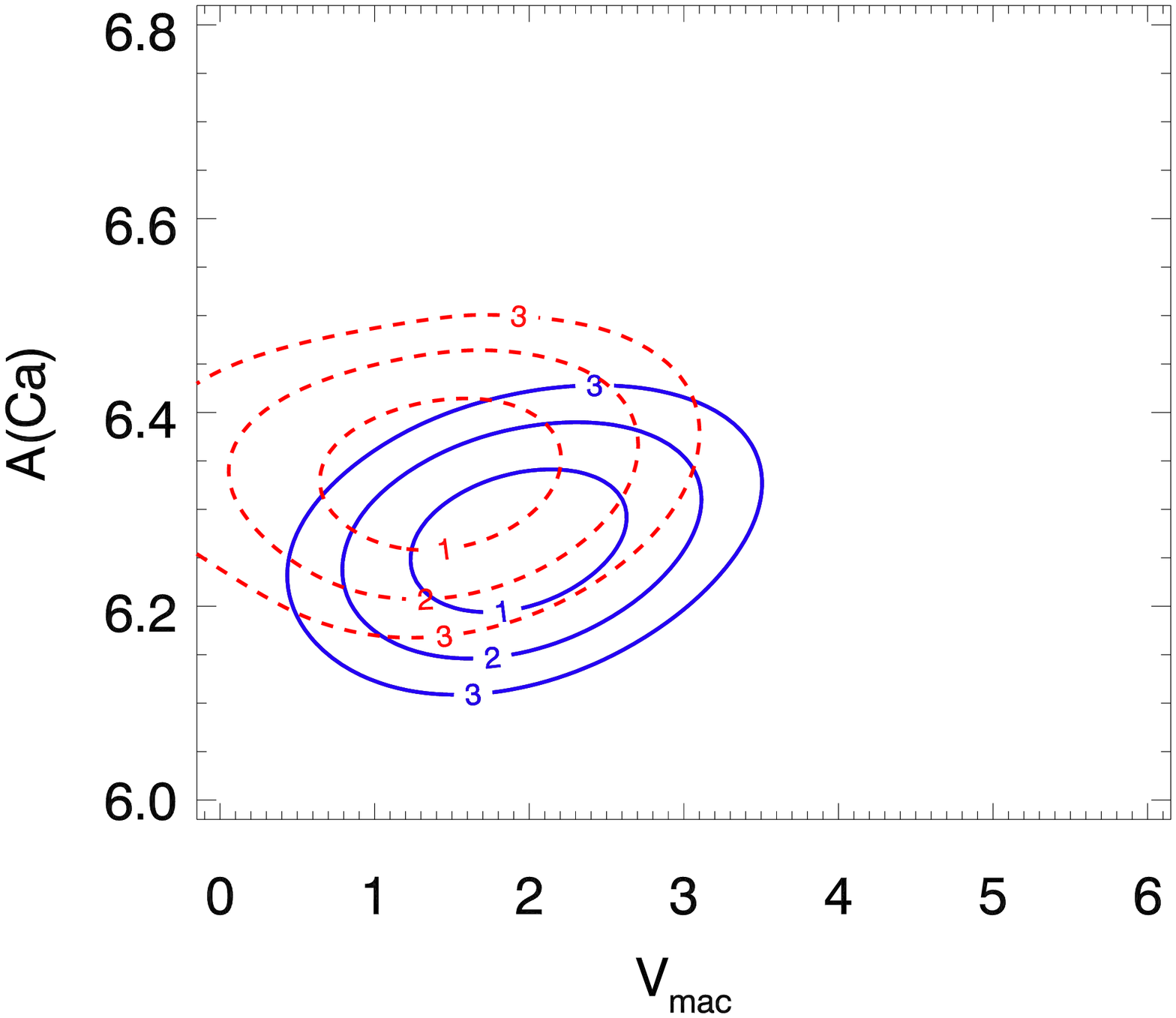}} 
      }; 
     \node[rotate=-90] at (0.30\textwidth,-0.16\textwidth) {Sun}; 
    \end{tikzpicture}  
    & 
    \hspace{-0.1\textwidth}\begin{tikzpicture} 
     \node[anchor=south west, inner sep=0,rotate=-90] (image) at (0,0) { 
      \subfloat{\includegraphics[width=0.45\textwidth]{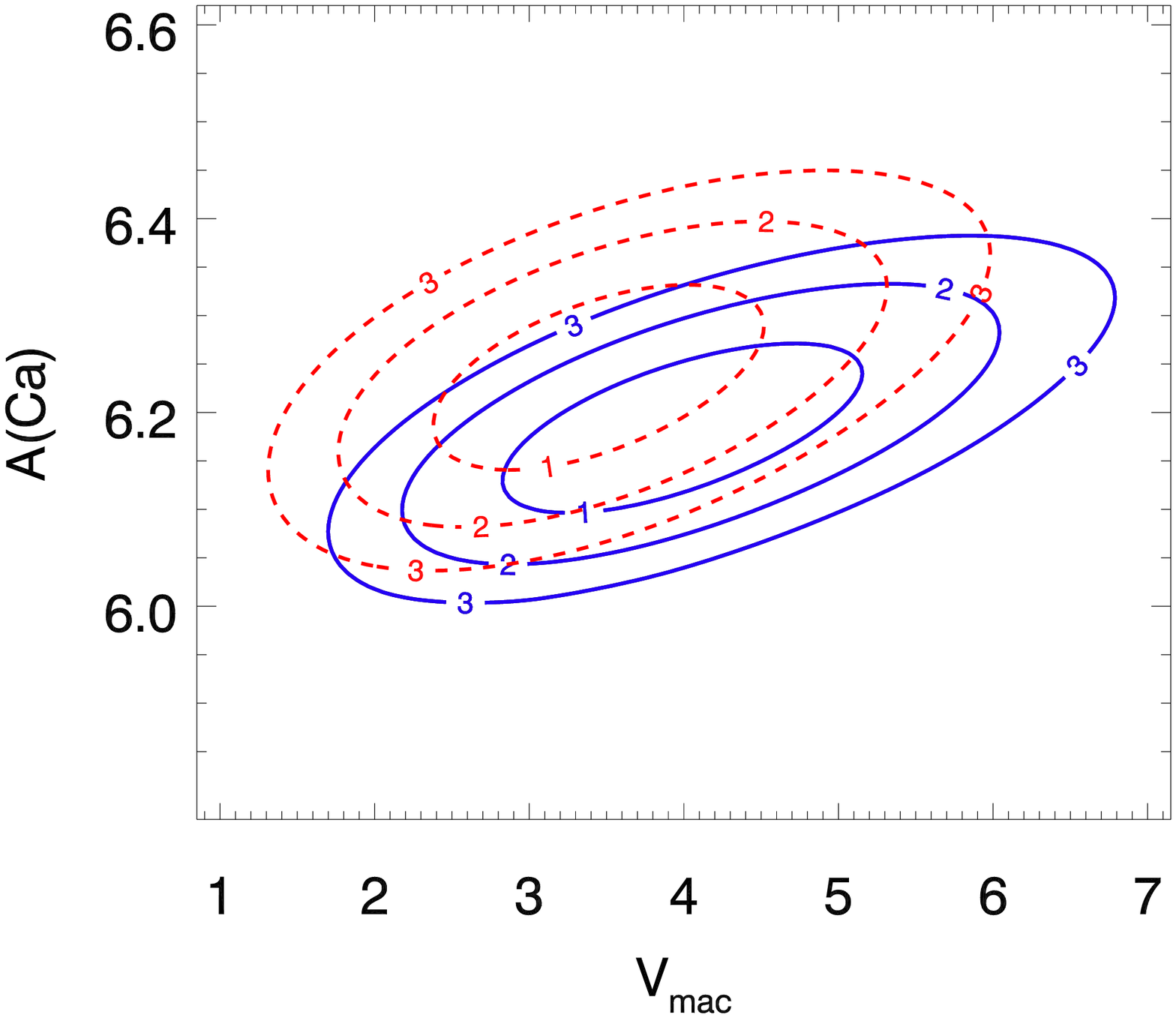}} 
      }; 
     \node[rotate=-90] at (0.30\textwidth,-0.16\textwidth) {Procyon}; 
    \end{tikzpicture} \\[-0.04\textwidth] 
   \end{tabular} 
\caption{Observed normalised flux spectra and comparison with our LTE / NLTE results. The spectra plots show the points used to calculate abundance and $V_{mac}$ and the residuals with best fit in LTE and NLTE. The contours show the confidence regions corresponding to 1, 2, and 3 $\sigma$ (see \citeauthor{1992nrfa.book.....P}, 1992).}\label{fig:allspec} 
\end{figure*}

Two methods were used here for the calculation of observed Ca abundances in each star. The first method is the so-called {\it line-by-line} method, where the abundance is calculated for individual lines. For each transition, the radial-tangential macro-turbulent velocity $v_{mac}$ and the Ca abundance were allowed to vary, in order to obtain the best possible match to the observations.  The dispersion in the inferred Ca abundances was used as an indicator of the uncertainty. In the second method, named {\it all-lines} method, the same Ca lines were analysed, but only one value of $v_{mac}$ and A(Ca) was allowed in order to find the best fit for {all} lines simultaneously. The best fit in each case was evaluated via $\chi^2$ minimisation; the error estimation for the all-lines method was determined via the confidence region in the [A(Ca),$v_{mac}$] plane where $\chi^2$ < min($\chi^2$)+2.71 (corresponding to a confidence level of 90\%), following the recommendations in \cite{1992nrfa.book.....P}. The confidence regions of the fit are shown in \fig{fig:allspec}.

\begin{figure*} 
\centering 
\begin{tabular}{ccc} 
 
\hspace{0.01\textwidth}\begin{tikzpicture} 
\node[anchor=south east, inner sep=0,rotate=-90] (image) at (0,0) {  
\subfloat{\includegraphics[width=0.70\textwidth]{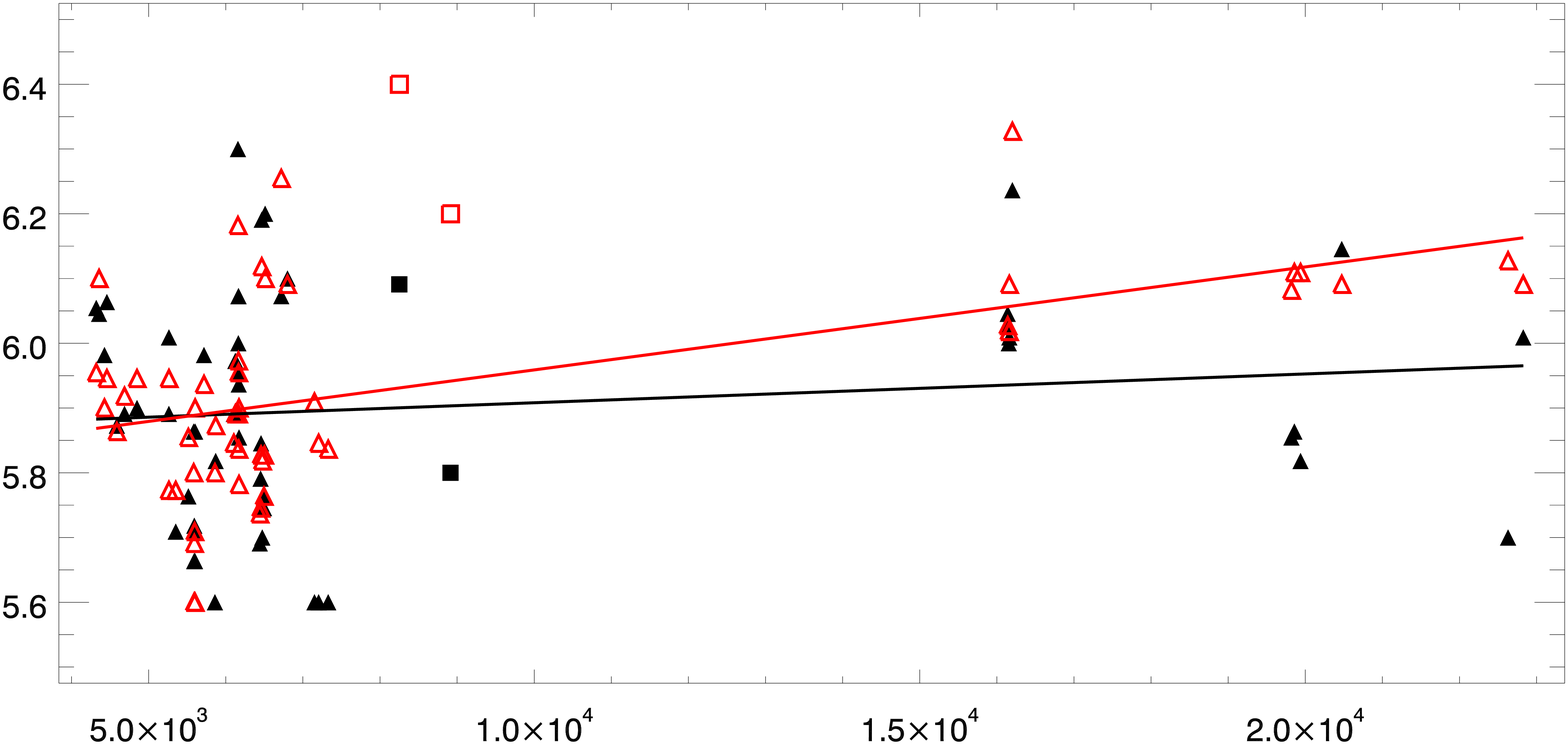}}};   
\node at (0.19\textwidth,0.65\textwidth) {\scalebox{1.00}{A(Ca)}};  
\node[rotate=-90] at (0.02\textwidth,0.32\textwidth) {\scalebox{1.0}{$\lambda$(\AA)}};  
\end{tikzpicture} 
& 
\hspace{-0.1\textwidth}\begin{tikzpicture} 
\node[anchor=south east, inner sep=0,rotate=-90] (image) at (0,0) {  
\subfloat{\includegraphics[width=0.70\textwidth]{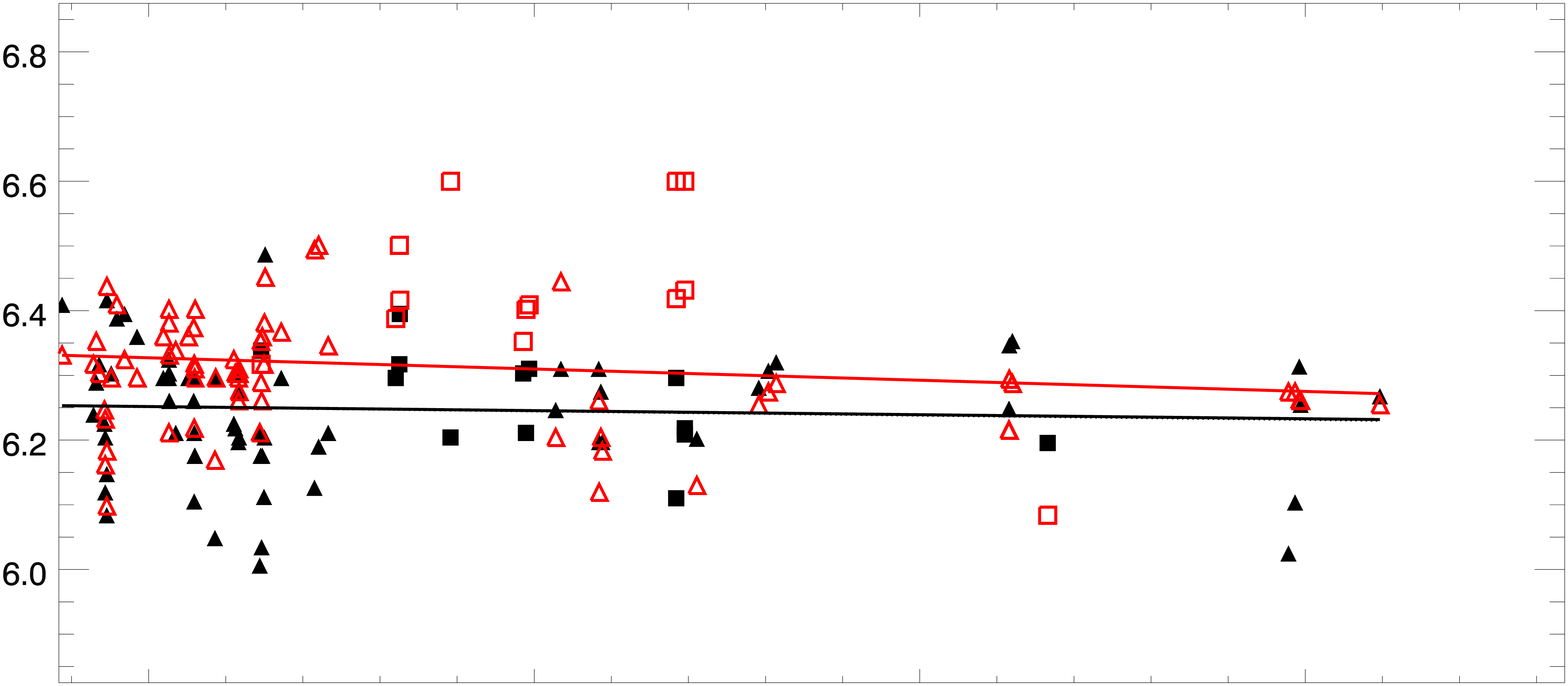}}};   
\node at (0.19\textwidth,0.65\textwidth) {\scalebox{1.0}{A(Ca)}};  
\end{tikzpicture}  
& 
\hspace{-0.1\textwidth}\begin{tikzpicture} 
\node[anchor=south east, inner sep=0,rotate=-90] (image) at (0,0) {  
\subfloat{\includegraphics[width=0.70\textwidth]{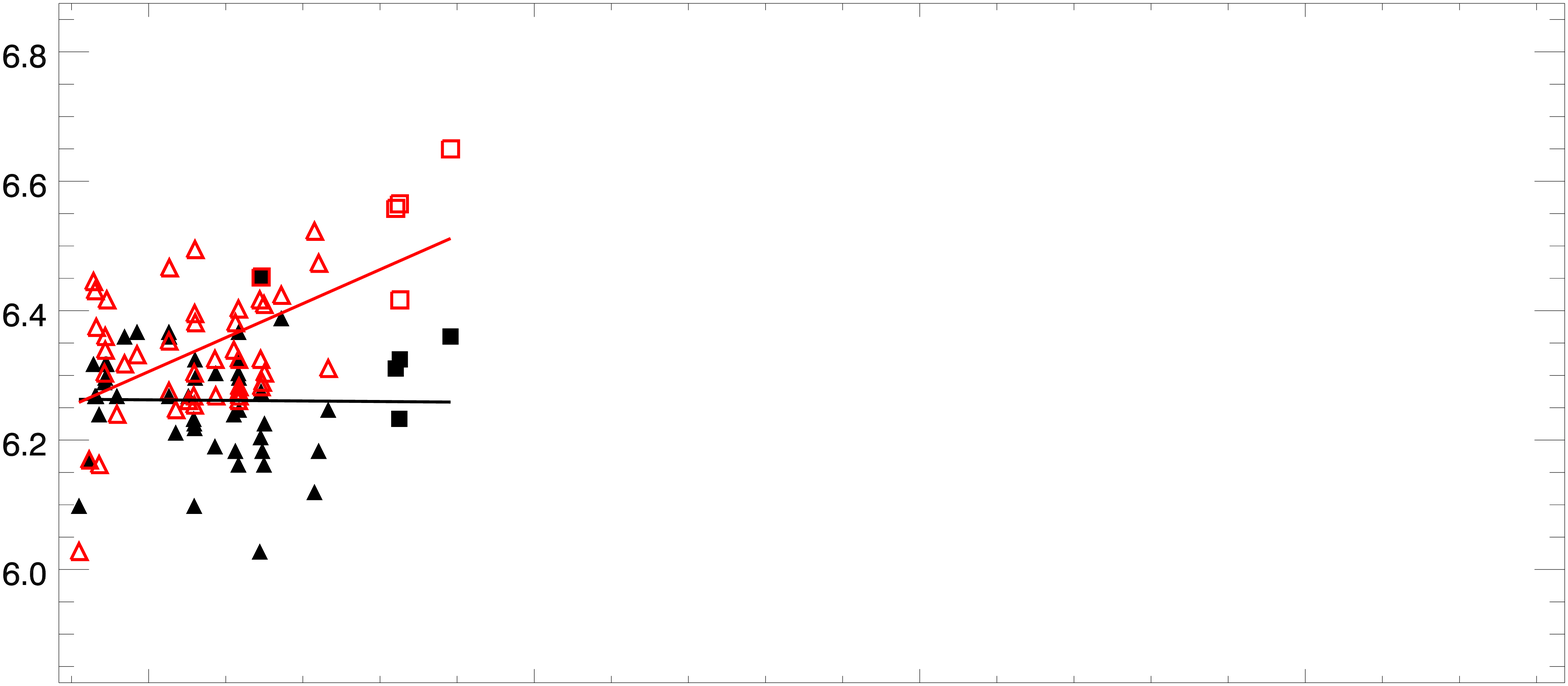}}}; 
\node at (0.19\textwidth,0.65\textwidth) {\scalebox{1.0}{A(Ca)}};   
\node[rotate=-90] at (0.14\textwidth,0.15\textwidth){\subfloat{\includegraphics[width=0.2\textwidth]{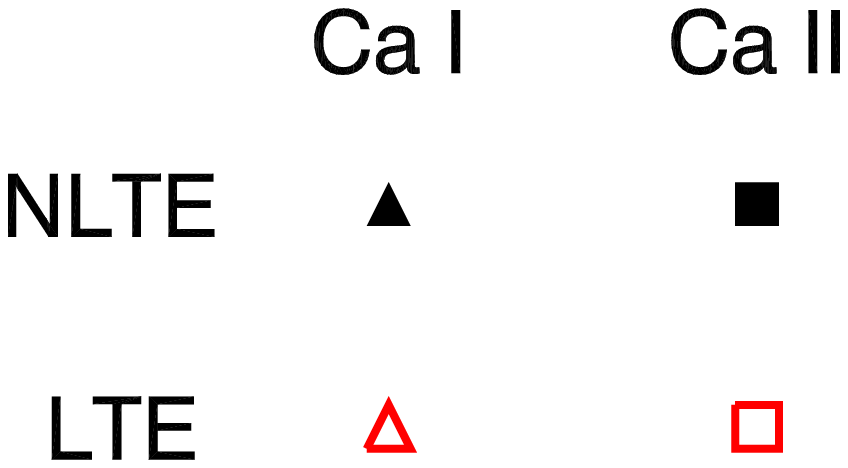}}}; 
\end{tikzpicture}\\[-0.1\textwidth]

\hspace{0.01\textwidth}\begin{tikzpicture} 
\node[anchor=south east, inner sep=0,rotate=-90] (image) at (0,0) {  
\subfloat{\includegraphics[width=0.70\textwidth]{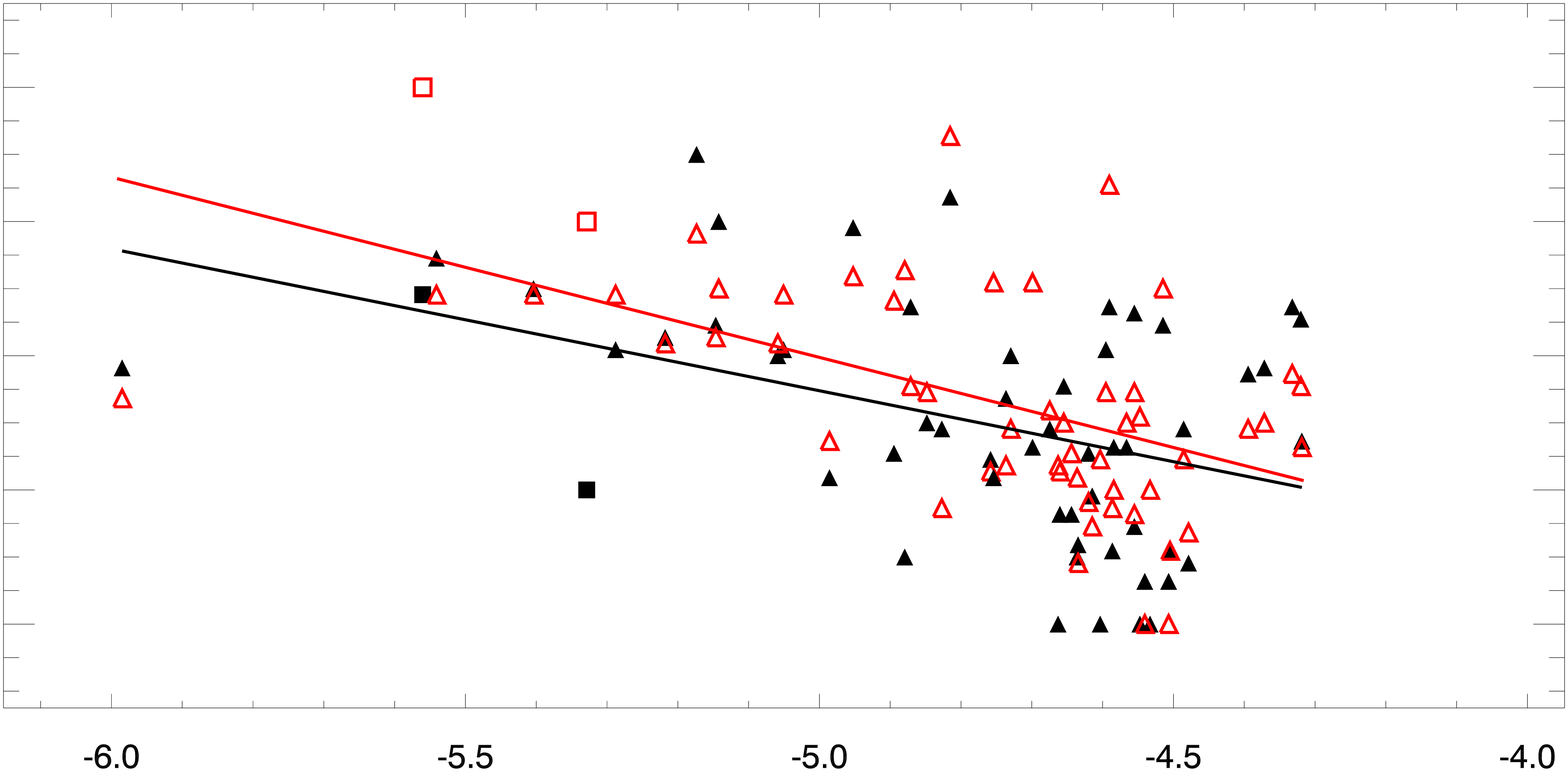}}};   
\node[right,rotate=-90] at (0.30\textwidth,0.6\textwidth) {\scalebox{1.3}{Arcturus}}; 
\node[rotate=-90] at (0.02\textwidth,0.32\textwidth) {\scalebox{1.0}{$\log_{10}$ (w$_{\rm{eq}}/\lambda$)}};  
\end{tikzpicture} 
& 
\hspace{-0.1\textwidth}\begin{tikzpicture} 
\node[anchor=south east, inner sep=0,rotate=-90] (image) at (0,0) {  
\subfloat{\includegraphics[width=0.70\textwidth]{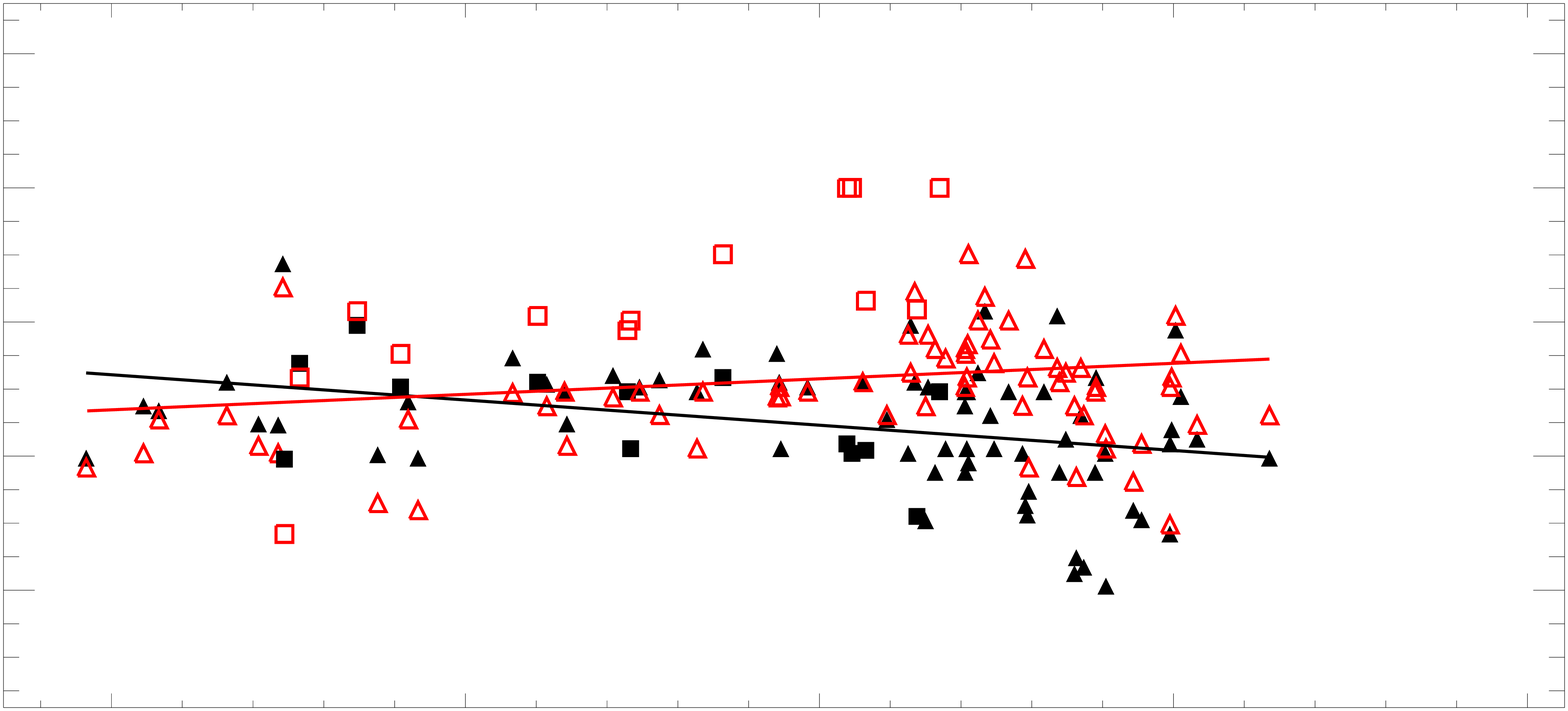}}};   
\node[right,rotate=-90] at (0.30\textwidth,0.6\textwidth) {\scalebox{1.3}{Sun}};   
\end{tikzpicture}  
& 
\hspace{-0.1\textwidth}\begin{tikzpicture} 
\node[anchor=south east, inner sep=0,rotate=-90] (image) at (0,0) {  
\subfloat{\includegraphics[width=0.70\textwidth]{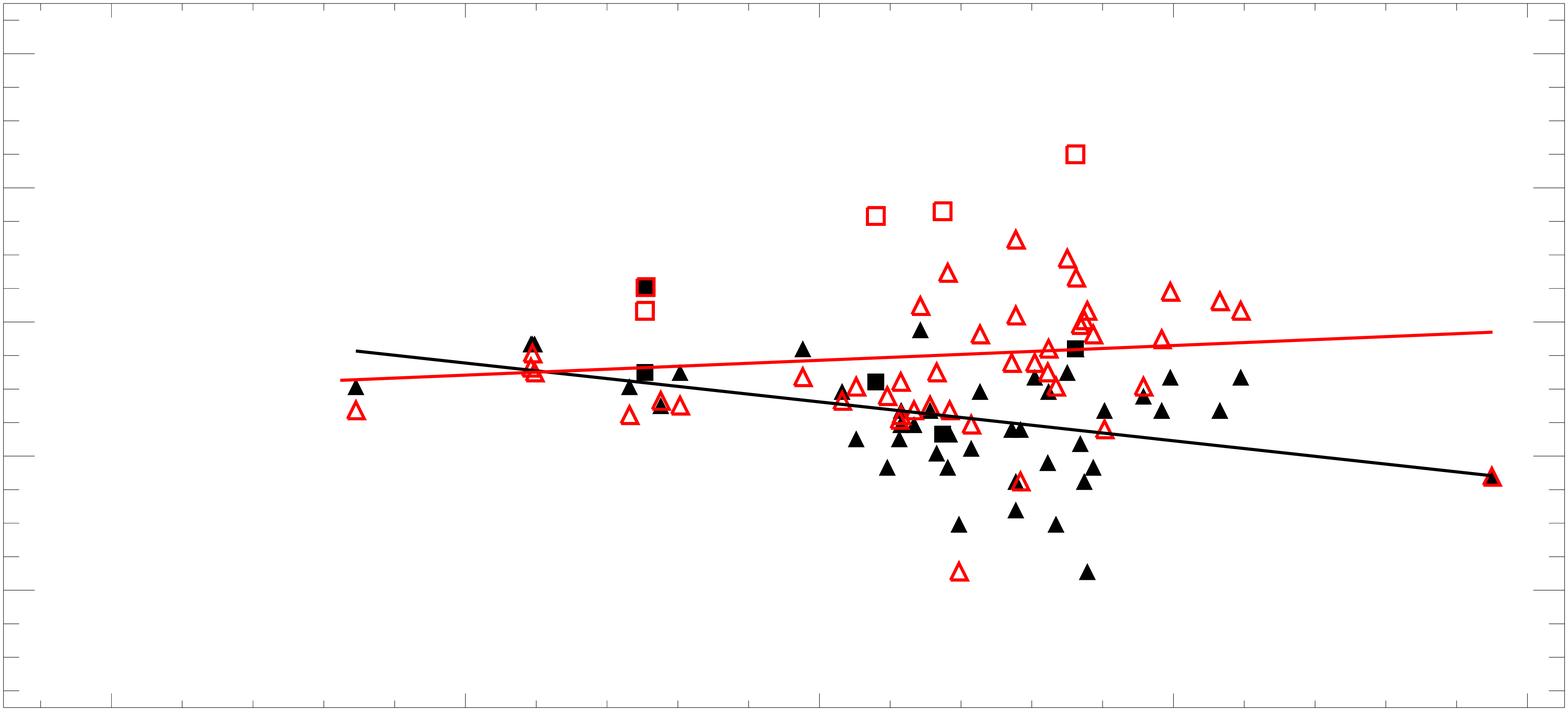}}};   
\node[right,rotate=-90] at (0.30\textwidth,0.6\textwidth) {\scalebox{1.3}{Procyon}};   
\end{tikzpicture}\\[-0.01\textwidth] 
 
\end{tabular} 
\caption{Line-by-line Ca abundance determination in the three observed stars. Red and black symbols denote LTE and NLTE abundances, respectively; triangles correspond to \cai\ and squares/asterisks to \caii\ lines. The lines show the linear fit of the LTE (red) and NLTE (black) abundances. }\label{fig:obslinebyline} 
 \end{figure*}

For the Sun and Arcturus, 76 and 59 lines, respectively, were used  for \mbox{3\,800 $<\lambda<$ 22\,900 \AA.} For Procyon, 49 lines were used in the range  \mbox{3\,800 - 9\,100 \AA}. Figure \ref{fig:obslinebyline} shows the behaviour of the derived LTE and NLTE abundances with respect to wavelength as well as line strength\footnote{Defined here as the unitless quantity w$_{eq}/\lambda$, where w$_{eq}$ and $\lambda$ are the equivalent width and wavelength of the line, respectively.}. We also experimented with different values of micro-turbulence that basically changed the slopes shown in \fig{fig:obslinebyline}. Tuning the micro-turbulence does not flatten the trends of \Aca{}{} versus line strength and \Aca{}{} versus wavelength simultaneously.

In Procyon, we found a NLTE abundance correction of \acorr$\sim-0.07$~dex in the two methods. The all-lines method shows a weak dependence of the best fit with macro-turbulence since the confidence region around the best $\chi^2$ is elongated in $v_{\rm{mac}}$ (see contours in \fig{fig:allspec}). The line-by-line method shows improvement in the trend of NLTE calcium abundances versus wavelength in comparison with LTE abundances but the trends of \Aca{}{} versus line strength show a negative slope in NLTE (see \fig{fig:obslinebyline}) while LTE derived abundances versus line strength show a positive slope.

\begin{figure}[ht] 
\hspace{0.0\textwidth} \begin{tikzpicture} 
    \node[anchor=south east, inner sep=0] (image) at (0,0) { 
      \subfloat{\includegraphics[width=0.48\textwidth]{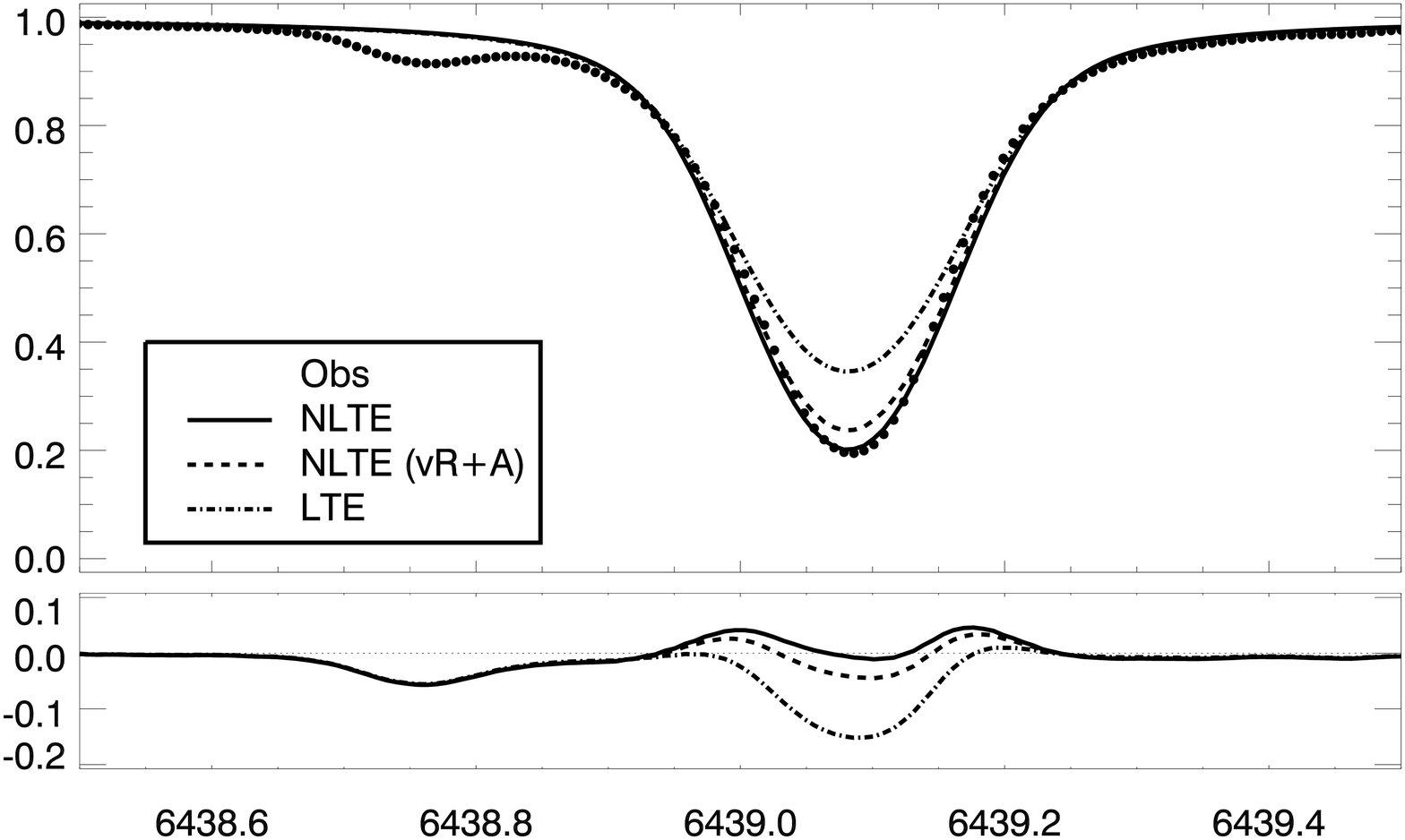}} 
    }; 
     \draw[black,fill=black] (-0.403\textwidth,0.16\textwidth) circle (0.002\textwidth);  
\end{tikzpicture} 
  \caption{Upper panel: Comparison between observations (solid circles) and line profiles in LTE (dot-dash), NLTE using vR+A electron collisional data (dot-dash), and NLTE with updated electron collisional data of the solar \cai\ 6439~\AA\ line [\Aca{}{}=6.25, V$_{\rm{mac}}$=3.0~m\,s$^{-1}$]. The low panel shows differences with observations.}\label{fig:solar-line} 
\end{figure}

In the solar spectrum, there is a small reduction in the spread in the derived \Aca{}{} in NLTE with respect to LTE. The standard deviations of the derived abundances in the line-by-line method change from 0.17 (IR) and 0.10 (visual) in LTE to 0.09 (IR and visual) in NLTE. Saturated lines ($-5.5\lesssim\log(w_{\rm{eq}}/\lambda)\lesssim-4.5$) suffer the strongest departures from LTE, while weak and strong lines are less affected (mid-right panel in \fig{fig:obslinebyline}). The strengthening of the line cores in NLTE significantly improves the match with observations. A typical example of the LTE and NLTE fits of a Ca line is shown in \fig{fig:solar-line}.  
As mentioned in  \cite{2017A&A...605A..53M}, we found also that NLTE effects in saturated lines are the \emph{weakening} of the wings and the \emph{strengthening} of the line core. The dependence of the NLTE abundance corrections with wavelength is less clear; lines with noticeable differences in LTE/NLTE abundance are spread all over the wavelength range (see left panels in \fig{fig:obslinebyline}). It is worth pointing out that the Ca lines covered by APOGEE \citep{apogee} at $\sim1.6\,\mu$m are not affected by NLTE in the Sun due to their small strength and formation depth.

\begin{table*}[htp] 
\caption{Parameters adopted and results of the calculation of inferred Ca abundance. The number of Ca lines used in each fit is in the sixth column ($n$). $\sigma$ represents the standard deviation of the quantity to the left. The errors in columns 12 and 13 were obtained based on the confidence levels obtained during the $\chi^2$ minimisation process.}\label{tab:observations} 
\centering                            
\begin{tabular}{lcccccrlllrlrr}\hline\hline 
                   &               &         &          &          &            &  &   \multicolumn{4}{c}{Line-by-Line}  & \multicolumn{3}{c}{All-lines}\\  
  Name     & \teff & \logg &     \feh         & v$_{mic}$  & $n$  & & \multicolumn{1}{c}{v$_{mac}$} & \multicolumn{1}{c}{$\sigma$} & \multicolumn{1}{c}{\small A(Ca)} & \multicolumn{1}{c}{$\sigma$} &  \multicolumn{1}{c}{v$_{mac}$}  & \multicolumn{1}{c}{\small A(Ca)} & \multicolumn{1}{c}{$\chi^2_{\rm{red}}$}\\ 
   & \small K  & \small cm\,s$^{-2}$ &  \small dex        & \small km s$^{-1}$  &    &  & \multicolumn{1}{c}{\small km s$^{-1}$} &   & \multicolumn{1}{c}{\small dex} &   &  \multicolumn{1}{c}{\small km s$^{-1}$}  & \multicolumn{1}{c}{\small dex} &      \\\hline\\[-0.3cm] 
Procyon  & $6\,530$ & $4.00$ & +0.0  & 2.00       &  49  & NLTE & 3.8 & 1.1 & 6.20 & 0.11 & $3.98\pm1.16$   & $6.18\pm0.09$ &   2.29 \\ 
          &          &        &      &            &      & LTE  & 3.2 & 1.0 & 6.27 & 0.10 & $3.38\pm1.07$   & $6.26\pm0.10$ &   3.92 \\ 
    Sun   & $5\,772$ & $4.44$ & +0.0 & 1.10       &  65  & NLTE & 2.0 & 0.4 & 6.25 & 0.09 & $1.97\pm0.69$   & $6.27\pm0.07$ &   3.19 \\ 
          &          &        &      &            &      & LTE  & 1.4 & 0.4 & 6.33 & 0.10 & $1.42\pm0.80$   & $6.34\pm0.09$ &   4.99 \\ 
   Sun(IR) &          &        &        &         &  16  & NLTE & 2.3 & 0.3 & 6.24 & 0.09 & $2.22\pm0.84$   & $6.24\pm0.06$ &   3.78 \\ 
         &          &        &      &             &      & LTE  & 1.5 & 0.4 & 6.26 & 0.17 & $1.49\pm0.81$   & $6.28\pm0.06$ &   4.90 \\ 
$\alpha$ Boo & $4\,247$ & $1.59$ & $-$0.5 & 1.63  &  44  & NLTE & 3.0 & 0.4 & 5.90 & 0.18 & $3.14\pm0.62$   & $6.01\pm0.09$ &   4.93 \\ 
         &          &        &      &             &      & LTE  & 2.9 & 0.4 & 5.90 & 0.13 & $2.82\pm0.62$   & $6.00\pm0.09$ &   3.71 \\ 
$\alpha$ Boo(IR) &       &       &         &      &  15  & NLTE & 3.4 & 0.4 & 6.08 & 0.30 & $3.48\pm0.75$   & $6.06\pm0.07$ &   1.46 \\ 
         &          &        &      &             &      & LTE  & 3.1 & 0.5 & 6.23 & 0.21 & $3.05\pm0.73$   & $6.18\pm0.08$ &   1.32 \\ 
\hline 
\end{tabular} 
\end{table*}%
 
\begin{figure*} 
\begin{tikzpicture} 
\hspace{0.0\textwidth} 
\node[anchor=south east, inner sep=0] (image) at (0,0) {  
  \subfloat{\includegraphics[width=0.98\textwidth]{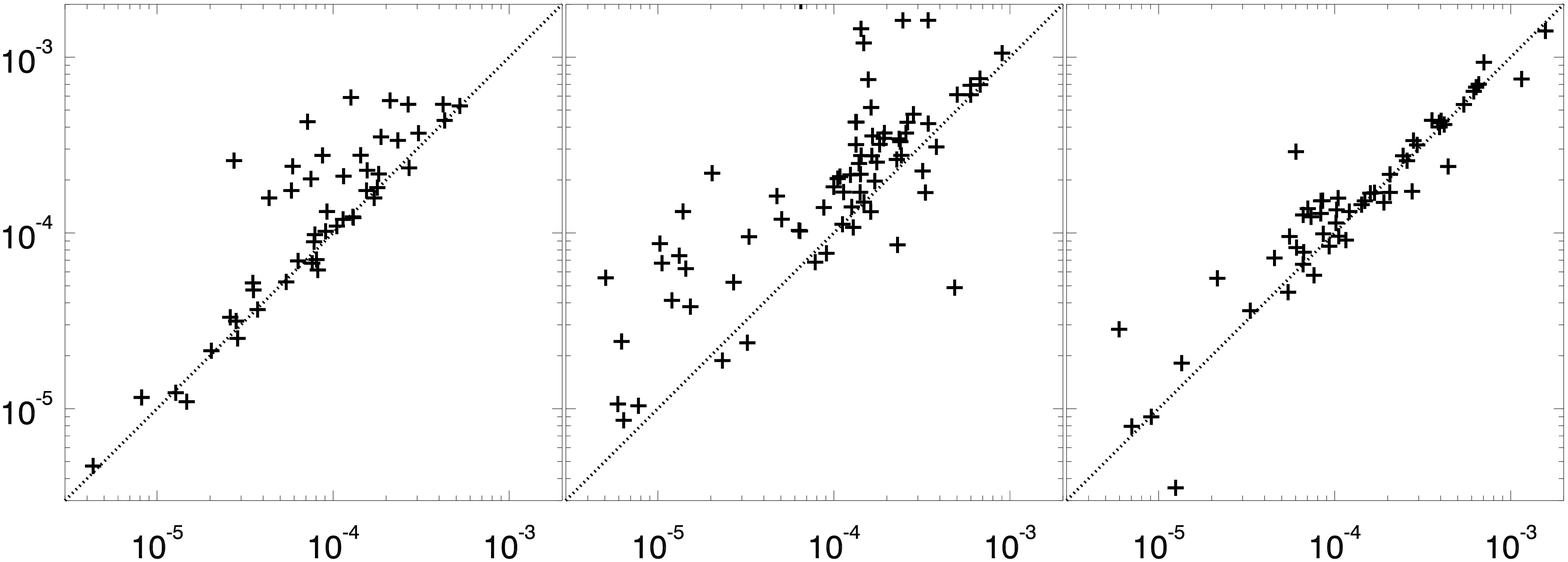}} 
}; 
\node[anchor=east] at (-0.64\textwidth,0.06\textwidth) {\scalebox{1.4}{Procyon}}; 
\node[anchor=east] at (-0.33\textwidth,0.06\textwidth) {\scalebox{1.4}{Sun}}; 
\node[anchor=east] at (-0.015\textwidth,0.06\textwidth) {\scalebox{1.4}{Arcturus}}; 
\node[rotate=90] at (-0.99\textwidth,0.20\textwidth) {\scalebox{1.1}{$<r>_{lte}$}}; 
\node[anchor=east] at (-0.74\textwidth,-0.01\textwidth) {\scalebox{1.1}{$<r>_{nlte}$}}; 
\node[anchor=east] at (-0.43\textwidth,-0.01\textwidth) {\scalebox{1.1}{$<r>_{nlte}$}}; 
\node[anchor=east] at (-0.11\textwidth,-0.01\textwidth) {\scalebox{1.1}{$<r>_{nlte}$}}; 
\end{tikzpicture}  
\caption{Comparison of the mean of residuals $<r>$ between observed and synthetic spectra after fitting each line separately (i.e. using the line-by-line method) between NLTE (abscissa) and LTE (ordinate) for the three stars studied here. The diagonal lines show the 1:1 relation between the two axes.}\label{fig:residuals} 
\end{figure*}

In the Sun, NLTE effects are obvious in the core of strong but unsaturated lines. The derived abundances obtained from line-profile fitting or equivalent-width matching are not significantly affected by this discrepancy, but the NLTE line profiles fit the observations much better, reducing the overall discrepancy between modelled and observed spectra (see \fig{fig:residuals}) . For example, in \fig{fig:solar-line} a $\chi^2$ test comparing the best fit and the observations leads to $\chi^2=$345.7, 191.3, and 160.6 for the LTE, NLTE (using the vR+A set of electron collisions), and NLTE cases, respectively\footnote{For this calculation we used 101 frequencies, two degrees of freedom, \Aca{}{} and v$_{mac}$, and an error of 0.01 (relative units) for the observed flux.}. We find consistent NLTE abundances for the visual and IR regions of the Sun (in the visual, \acorr$\sim -0.08$~dex). We also note that the derived macro-turbulent velocity is lower in LTE than in NLTE.  The resulting NLTE correction is also a consequence of having v$_{\rm{mac}}$ as a free parameter; in order to reduce the overall difference between observed and synthetic line profiles, v$_{\rm{mac}}$ tends to decrease when \Aca{LTE}{} increases.

In Arcturus, NLTE improves the agreement between derived abundances in the visual and the IR. When LTE is adopted, the difference of derived \Aca{}{} between the visual and the IR is 0.18~dex when using the all-lines method and 0.33~dex when using the line-by-line method. The derived \Aca{NLTE}{} in the visual and IR regions differ by 0.05~dex when using the all-lines method. The line-by-line method shows a discrepancy of \Aca{NLTE}{} between the visual and the IR of 0.18~dex but there is no improvement in the dispersion, with respect to the LTE case, and marginal improvements in the trends with $\lambda$ and line strength (see \fig{fig:obslinebyline}). Improving the agreement of derived abundances in different wavelength regimes is especially important for studies that aim to compare abundances from different surveys \citep{2018arXiv180707625A}.

For the three cases studied here, we checked the effect of microturbulence in the derived NLTE abundances and did not find any significant improvement either in the overall fit obtained or in the dispersion of derived abundances in the line-by-line method. The trends in wavelength and line strength (see \fig{fig:obslinebyline}) observed when using the line-by-line method cannot be corrected simultaneously by tuning the microturbulent velocity.

The residuals between the synthetic and observed line profiles are improved in NLTE with respect to LTE in the three cases studied here. In \fig{fig:residuals} we compared the mean of the residuals \footnote{defined for line $i$ as \[<r_i>=\frac{1}{N_i}\sum_{j=1}^{N_i}(O_{ij}-S_{ij})^2\] where $N_i$ is the number of points used to fit line $i$. $O$ and $S$ stand for observed and synthetic points respectively.} of the best fit obtained for each line in LTE and NLTE using the line-by-line method. We can conclude that in general NLTE reproduces the line profiles better than LTE.

Regarding the \cai/\caii\ abundance ratio, we obtain good agreement in NLTE (see Table \ref{tab:caratioalllines}). When applying the all-lines method in the solar IR spectra we used 13 \cai\ lines, and only the two \caii\ lines at $\sim$11\,900 \AA\ because the other two \caii\ lines covered by the observed spectrum were too weak.  
When using the line-by-line method (grouping together all the lines in the visual and IR solar spectra, see Table \ref{tab:caratioalllines}) we again obtain a good agreement in the NLTE results. We obtained very similar 1D NLTE results to \cite{2015A&A...573A..25S}, who used MARCS models (6.26 and 6.28~dex for \cai\ and \caii\ respectively). However, they recommend \Aca{\sun}{}=6.32$\pm$0.03~dex resulting from their 3D radiative-transfer calculations combined with $\langle 3D\rangle$ NLTE corrections. 
 
The work of \cite{2017A&A...605A..53M} presents a reduction of the \cai/\caii\ discrepancy $\Delta A(\caii,\cai)$ from  0.13~dex in LTE to 0.07~dex in NLTE for the Sun. They also derived A(Ca) from the 8498 \AA\ \caii\ line, which is insensitive to NLTE effects in the Sun and with a value of \Aca{8498}{} = 6.27~dex is in agreement with our NLTE results (see Table \ref{tab:caratioalllines}). 
 
In the case of Procyon, \citeauthor{2017A&A...605A..53M} found good agreement between the derived \cai\ and \caii\ abundances already in LTE (0.03~dex) and the discrepancy slightly increases in NLTE (0.06~dex). In Table \ref{tab:caratioalllines} we show our results (in LTE $\sim0.25$~dex and NLTE $\sim0.04$~dex). The reason for this difference could be due to the fact the \caii\ lines used to report the \caii\ abundances in column 3 of Table \ref{tab:caratioalllines} do not include the \caii\ triplet lines (which for Procyon and Arcturus are listed in column 5 of the same table).

\begin{table*} 
  \centering 
  \caption{Derived LTE/NLTE abundances from \cai\ and \caii\ lines using the all-lines and the line-by-line methods for each star. In front of the names of the stars is the number of lines used. For the all-lines method, the values in parenthesis represent the error. Values with asterisks were obtained using only the wings of the \caii\ IR triplet lines (see text). For the line-by-line method, $\sigma$ represents the standard deviation of the quantity to the left.}\label{tab:caratioalllines} 
  \begin{tabular}{ r c c c c c c c c c}\hline\hline 
    &    \multicolumn{4}{c}{All-lines}  & \multicolumn{4}{|c}{Line-by-Line}\\ 
    & \multicolumn{2}{c}{Visual}& \multicolumn{2}{c}{IR} \\  
           & \cai &    \caii  &     \cai  &  \caii     &     A(\cai) & $\sigma$  & A(\caii)  &  $\sigma$  \\\hline 
\multicolumn{1}{l}{Procyon} &  44  &   5       &           &            &        44    &         &     5     \\                                         
NLTE   & 6.19(9)  &  6.22(8)   &           &  6.22(4)*  &       6.20   &  0.10   &   6.24    &    0.11      \\                       
LTE    & 6.21(10) &  6.50(9)   &           &  6.21(4)*  &       6.24   &  0.11   &   6.49    &    0.10      \\\hline            
\multicolumn{1}{l}{Sun}&  55   &   10      &   13       &   3        &        68    &          &    13     \\                                 
NLTE   & 6.26(10) &  6.24(6)   & 6.25(12)  &  6.20(10)  &       6.24   &  0.10   &   6.26    &    0.09      \\            
LTE    & 6.25(10) &  6.32(12)  & 6.29(14)  &  6.44(8)   &       6.30   &  0.09   &   6.42    &    0.13      \\\hline       
\multicolumn{1}{l}{Arcturus\phantom{xxx}} &  42 &     2   &  15     &    &        57    &        &     2    \\                            
NLTE   & 6.02(21) &  6.04(9)   & 6.11(15)  &  6.04(8)*  &       5.97   &  0.14   &   6.04    &           \\ 
LTE    & 6.00(20) &  6.38(16)  & 6.21(15)  &  6.05(8)*  &       6.00   &  0.13   &   6.38    &            \\\hline   
\end{tabular} 
\end{table*}

In the three stars studied here, the wings of the \caii\ IR lines are not sensitive to either NLTE or macro-turbulence. In the Sun, the cores of these lines in LTE have depths similar to the observed ones but once NLTE is included the synthetic lines get deeper. When the solar model atmosphere includes chromosphere, the NLTE cores fit the observations, while the LTE cores go into emission (see \fig{fig:clv_spec}).

\subsection{Centre-to-limb variation of solar Ca lines} 
 
Observations of the CLV of solar Ca lines were performed at the Swedish Solar Telescope (SST) in the summer of 2011. These data were used by \cite{2017MNRAS.468.4311L} who gave a detailed description of their analysis. In short, observations were performed using the TRIPPEL spectrograph \citep{2011A&A...535A..14K}, at $\mu={0.2,0.4,0.6,0.8}$ and $1.0$, covering several wavelength regions with spectral resolution of $\sim$~150\,000, among them the region used in this work: (6147,6159) \AA;  the \caii\ IR triplet line at 8662~\AA\ was also part of the observations.

\begin{figure*} 
  \centering 
   \begin{tabular}{c c} 
    \hspace{-0.09\textwidth}\begin{tikzpicture} 
      \node[anchor=south east, inner sep=0] (image) at (0,0) { 
        \subfloat{\includegraphics[width=0.6\textwidth]{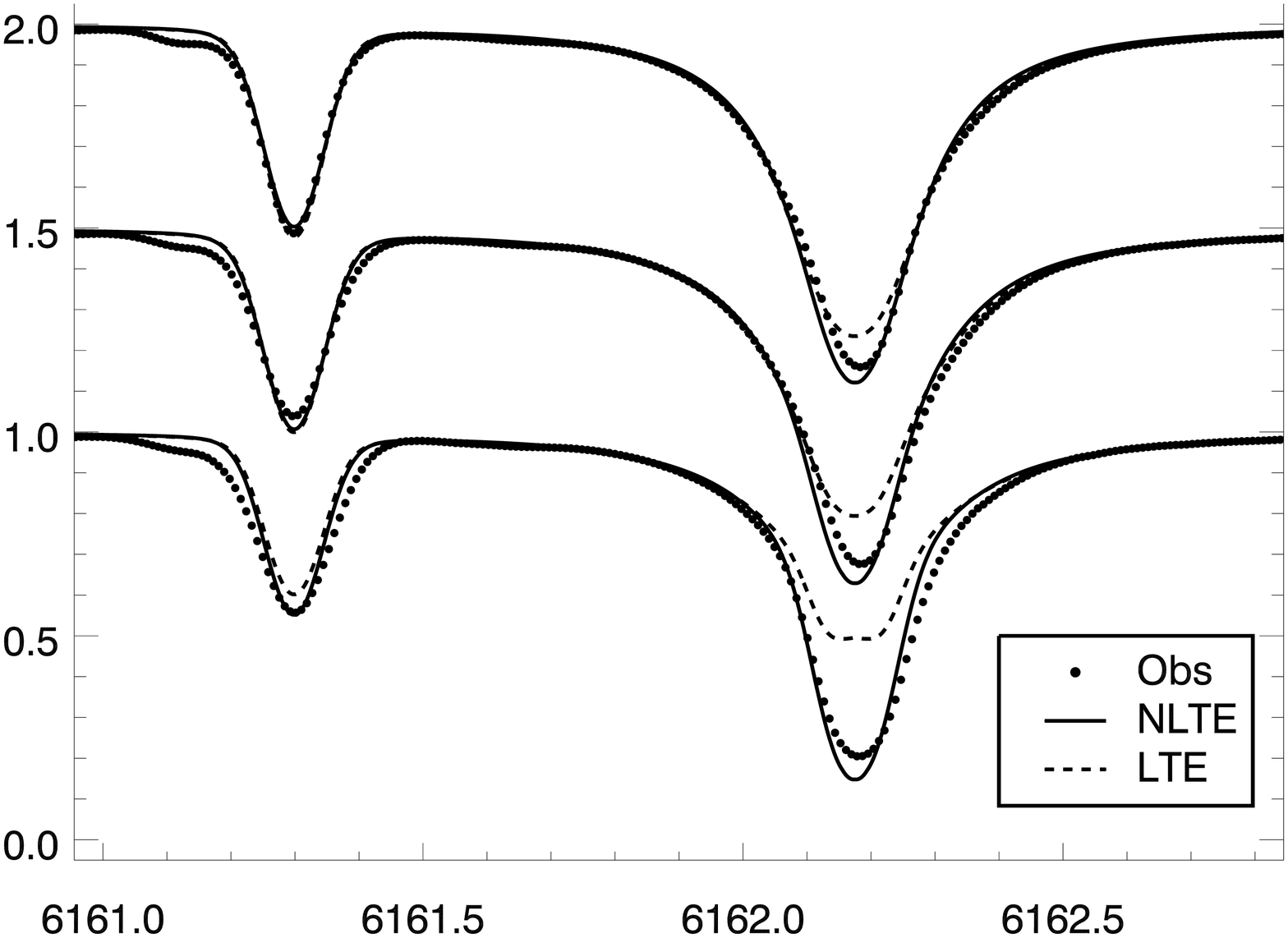}} 
      }; 
      \node at (-0.455\textwidth,0.395\textwidth) {\scalebox{0.8}{$\mu=1.0$}}; 
      \node at (-0.455\textwidth,0.320\textwidth) {\scalebox{0.8}{$\mu=0.6$}}; 
      \node at (-0.455\textwidth,0.243\textwidth) {\scalebox{0.8}{$\mu=0.2$}}; 
      \node[rotate=90] at (-0.52\textwidth,0.23\textwidth) {relative intensity}; 
    \end{tikzpicture} 
    & 
    \hspace{-0.12\textwidth}\begin{tikzpicture} 
     \node[anchor=south east, inner sep=0] (image) at (0,0) { 
      \subfloat{\includegraphics[width=0.6\textwidth]{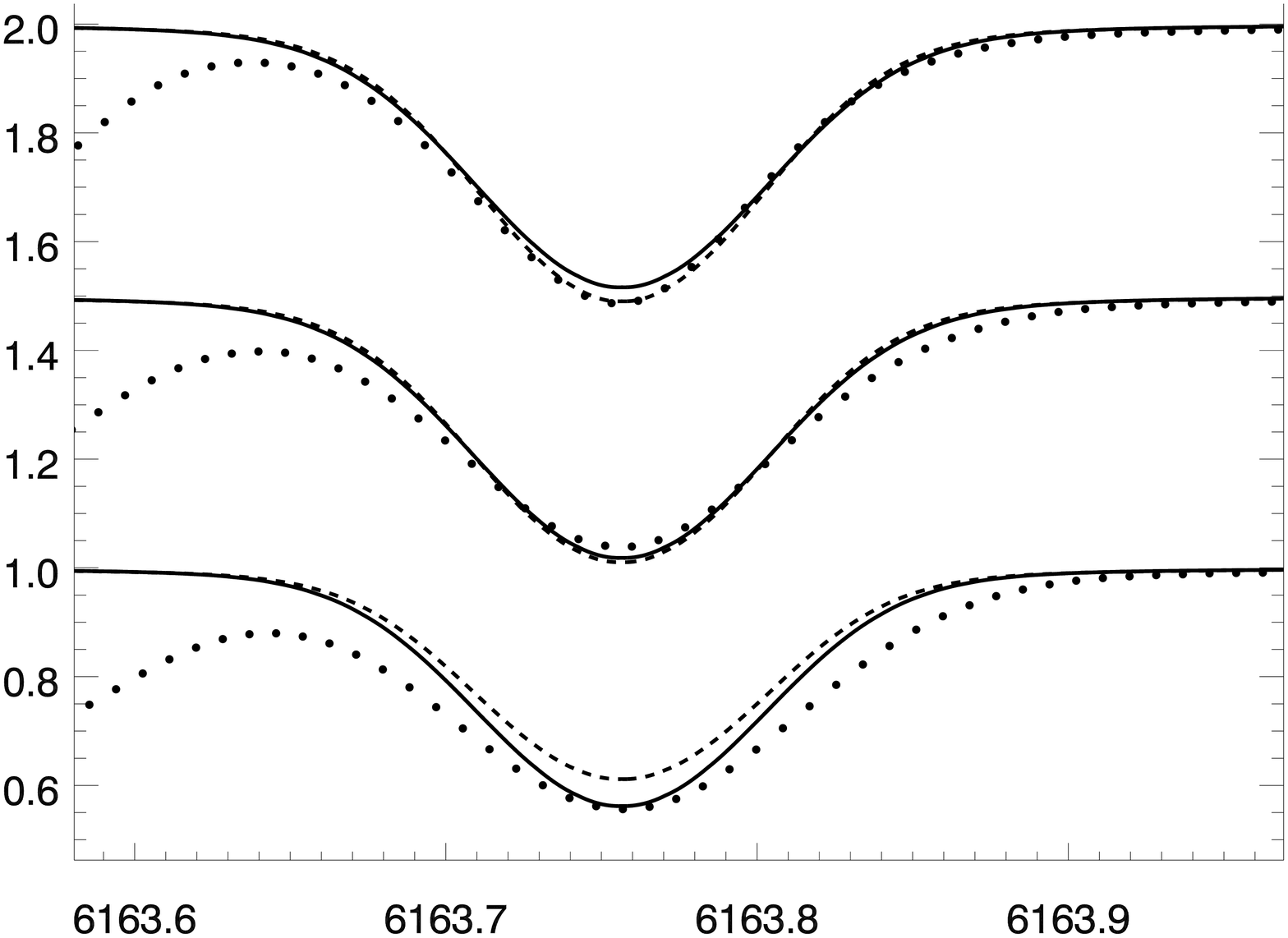}} 
      }; 
      \node at (-0.455\textwidth,0.395\textwidth) {\scalebox{0.8}{$\mu=1.0$}}; 
      \node at (-0.455\textwidth,0.293\textwidth) {\scalebox{0.8}{$\mu=0.6$}}; 
      \node at (-0.455\textwidth,0.193\textwidth) {\scalebox{0.8}{$\mu=0.2$}}; 
    \end{tikzpicture} \\[-0.08\textwidth] 
        \hspace{-0.09\textwidth}\begin{tikzpicture} 
      \node[anchor=south east, inner sep=0] (image) at (0,0) { 
        \subfloat{\includegraphics[width=0.6\textwidth]{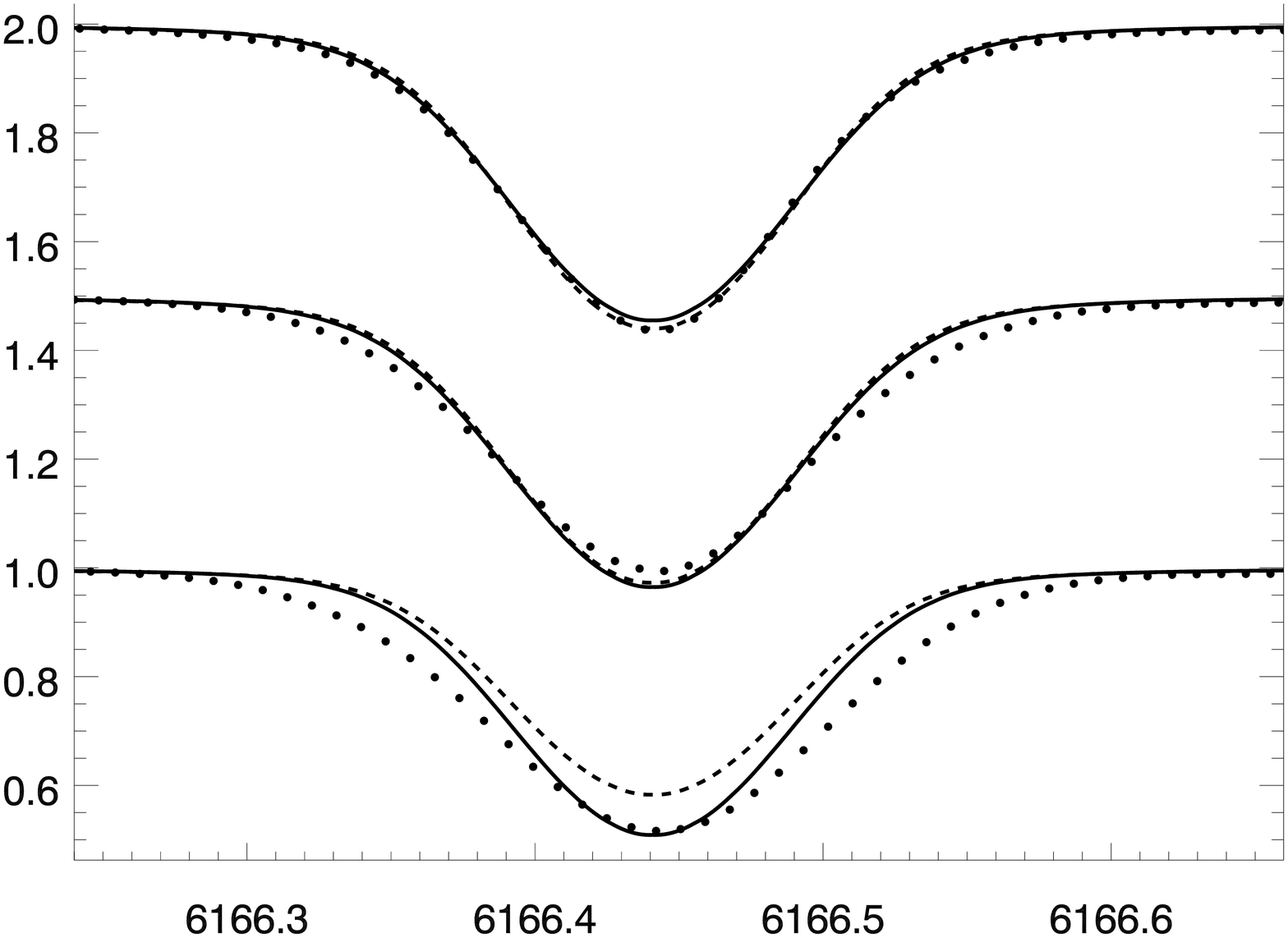}} 
      }; 
      \node at (-0.455\textwidth,0.395\textwidth) {\scalebox{0.8}{$\mu=1.0$}}; 
      \node at (-0.455\textwidth,0.293\textwidth) {\scalebox{0.8}{$\mu=0.6$}}; 
      \node at (-0.455\textwidth,0.193\textwidth) {\scalebox{0.8}{$\mu=0.2$}}; 
      \node[rotate=90] at (-0.52\textwidth,0.23\textwidth) {relative intensity}; 
    \end{tikzpicture} 
    & 
    \hspace{-0.12\textwidth}\begin{tikzpicture} 
     \node[anchor=south east, inner sep=0] (image) at (0,0) { 
      \subfloat{\includegraphics[width=0.6\textwidth]{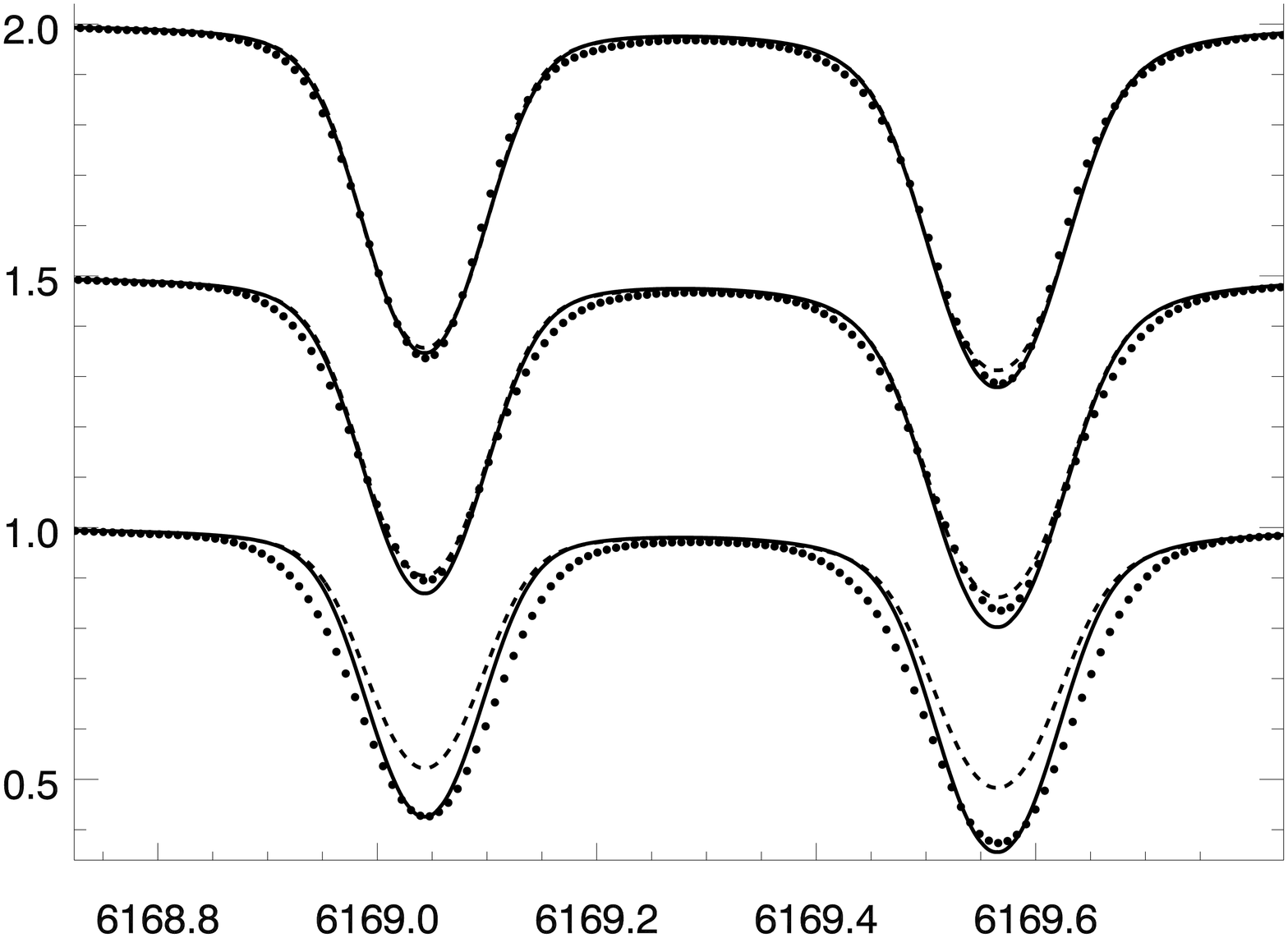}} 
      }; 
      \node at (-0.06\textwidth,0.394\textwidth) {\scalebox{0.8}{$\mu=1.0$}}; 
      \node at (-0.06\textwidth,0.299\textwidth) {\scalebox{0.8}{$\mu=0.6$}}; 
      \node at (-0.06\textwidth,0.205\textwidth) {\scalebox{0.8}{$\mu=0.2$}}; 
    \end{tikzpicture} \\[-0.08\textwidth] 
\multicolumn{2}{r}{ 
   \hspace{-0.12\textwidth}\begin{tikzpicture} 
     \node[anchor=south east, inner sep=0] (image) at (0,0) { 
      \subfloat{\includegraphics[width=1.1\textwidth]{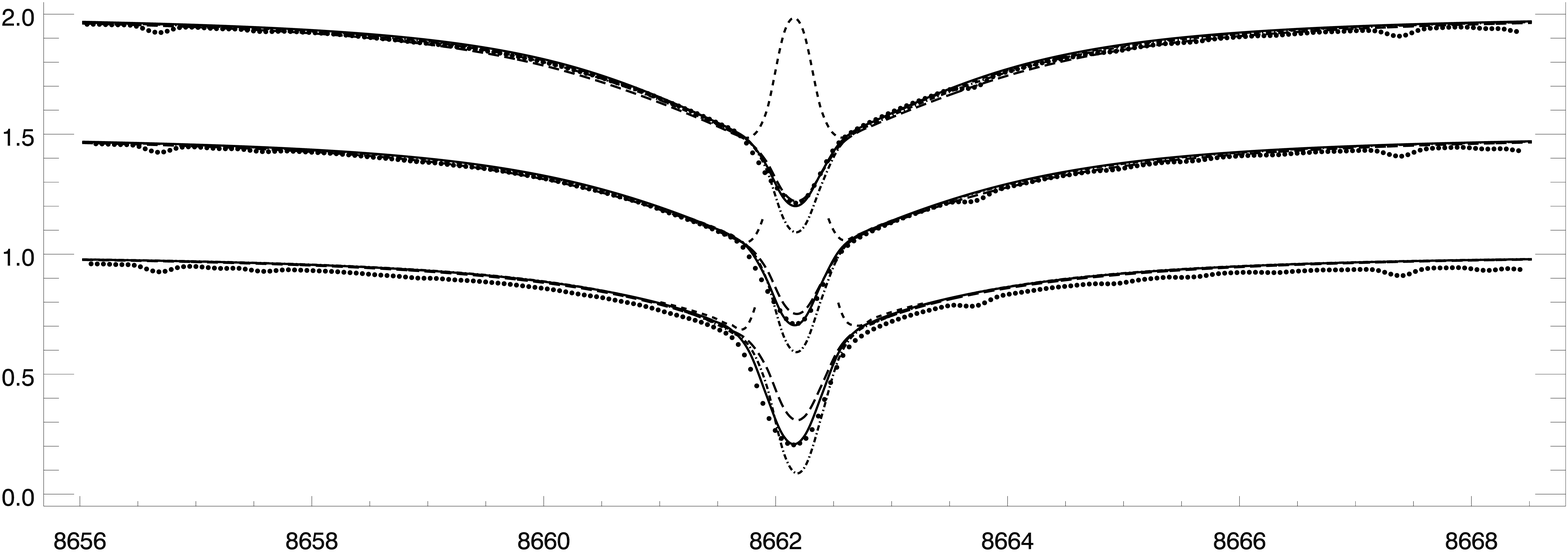}} 
      }; 
      \node at (-0.9\textwidth,0.385\textwidth) {\scalebox{0.8}{$\mu=1.0$}}; 
      \node at (-0.9\textwidth,0.310\textwidth) {\scalebox{0.8}{$\mu=0.6$}};  
      \node at (-0.9\textwidth,0.235\textwidth) {\scalebox{0.8}{$\mu=0.2$}}; 
      \node at (-0.5\textwidth,0.02\textwidth) {$\lambda$ (\AA)}; 
    \end{tikzpicture} 
    } 
  \end{tabular} 
\caption{Centre-to-limb variation of the Ca lines in the [6\,160,6\,170]~\AA\ region and the 8662~\AA\ \caii\ triplet line. The atmospheric model used was MACKKL. In order to get the best fit for the observed intensity profile at $\mu=1.0$, calcium abundances and macro-turbulent velocities were [A(Ca), $v_{macro}$]=(6.28~dex, 1.6~m\,s$^{-1}$) for the LTE and (6.26~dex, 1.9~m\,s$^{-1}$) for the NLTE synthetic profiles. The 8662~\AA\ line produces an emission core in LTE due to the increase in temperature at $\tau\sim-4$ in the MACKKL model; in order to avoid over-plotting, the emission cores at $\mu=$ 0.6 and 0.2 were cut. The \caii\ triplet line was also modelled using the Kurucz solar model atmosphere. LTE-Kurucz profiles are the long dashed lines and NLTE-Kurucz profiles are the dot-dashed lines.}\label{fig:clv_spec} 
\end{figure*} 
 
The CLV observations are shown in \fig{fig:clv_spec}. The synthetic spectra show only the Ca lines under study. The spectrum at $\mu=1.0$ was used to set A(Ca) and the macro-turbulent velocity using the all-lines method. We obtained [\Aca{}{}~dex,v$_{mac}$~ms$^{-1}$] = [6.26(5),~1.9(3)] in NLTE and [6.28(5),~1.6(3)] in LTE. The values found at $\mu=1.0$ were used in the synthetic line profiles at $\mu=$0.8, 0.6, 0.4 and 0.2. Figure \ref{fig:clv_spec} shows the comparison between the observations and the synthetic spectra for the LTE (dashed line) and NLTE (solid line) cases. The MACKKL solar model is used in this comparison. NLTE profiles are able to reproduce the depth of these lines at every $\mu$ value, while LTE profiles produce weaker cores at lower $\mu$. As seen in \fig{fig:clv_spec}, the depth of the NLTE cores match the observations even at low $\mu$ but the observed wings are wider than the synthetic ones. We believe the missing ingredient is atmospheric inhomogeneities; as shown by \cite{2017MNRAS.468.4311L} for the case of Fe, the observed variation of $w_{\rm{eq}}$ with $\mu$ is modelled significantly better with 3D than with 1D modelling.  
A more sophisticated analysis involving 3D model atmospheres and radiative transfer is expected to improve the agreement with observations near the limb. 
 
We also tested the CLV behaviour using the Kurucz solar model atmosphere and found that the LTE/NLTE behaviour is similar in both cases for five of the lines in the 6165~\AA\ region because those lines are formed in deep atmospheric layers. The 6162~\AA\ line, on the other hand, has a different profile in LTE when using the MACKKL and Kurucz solar atmospheres but the NLTE line profiles are almost identical (the line has the same depth in both cases but the core of the MACKKL profile is thinner). This is unexpected given that lines tend to form at higher layers in NLTE and this is where the two atmospheric models differ the most. We found, though, that different NLTE mechanisms are at work in the high layers of these two atmospheres: in the MACKKL model, the Planck function $B_\nu(T)$  increases from $\log$(\taufive)$ < -4$ making $S_v<B_\nu(T)$ and over-recombination of \cai\ levels from the \caii\ reservoir dominates. For the solar models with no chromosphere, levels of CaI tend to lower the population with respect to LTE (due to overall over-ionisation), but for the lowest levels, photon losses (particularly due to frequency redistribution from line wings) compete efficiently against over-ionisation in at high layers of the atmosphere making deeper line cores with respect to LTE, as in the case of the \cai\ lines at 6165~\AA.       
 
The core of the 8662~\AA\ \caii\ triplet line forms at higher layers than the \cai\ lines at $\sim$6165~\AA\ making the LTE-MACKKL line profile produce an emission core while the NLTE-MACKKL line profile matches the observations. The solar Kurucz model does not include the chromosphere. If this model is used, the wings of this line are almost unaffected, but the core differs considerably: at $\mu=1.0$ the LTE-Kurucz profile fits the observations better than the NLTE-Kurucz profile, but, as was the case for the 6162~\AA\ line, the former becomes too weak at lower $\mu$ (see bottom panel in \fig{fig:clv_spec}, where the long-dashed line shows the LTE-Kurucz profile). The profile for the NLTE-Kurucz model has, at every $\mu,$ deeper cores than the observations.

\section{Conclusions} 
 
We performed NLTE calculations of neutral and single-ionised calcium using state-of-the-art radiative and collisional data and improved approximation formulas for the missing collisional data. The results were tested against high-quality observations in a wide wavelength range on three benchmark stars: Procyon, Arcturus, and the Sun. Very recently \cite{2018ApJ...867...87B} presented detailed calculations on \caii+H collisions. Unfortunately that work was too recent to be considered here but will certainly be considered in our future investigations on NLTE effects of calcium. Our analyses have lead us to the following main conclusions. 
 
\begin{itemize} 
 
\item In Procyon, NLTE abundances are more consistent along the observed wavelength range used in this work than LTE. In the Sun, the dispersion in abundances derived for each line is slightly reduced in NLTE when compared to LTE. Removing outliers does not change the trends of derived abundances versus wavelength/line strength. In Arcturus, the average LTE abundances in the visual and the IR differ by 0.2~dex, which is reflected in a trend of derived abundance with wavelength; this trend is reduced in NLTE.  The main NLTE effect in the three stars studied here is visible in the core of lines of medium strength ($\rm{W}_{\rm{eq}}/\lambda\sim10^{-4}$).

\item The lowest states of \cai\ tend to be underpopulated with respect to the LTE values in the metal-poor giant Arcturus while for the Sun and Procyon, low-lying levels of \cai\ are overpopulated with respect to LTE at line-formation depths. Due to its weak electron collisional coupling (obtained from quantum mechanical calculations), the 3d$^2$($^3$F) level suffers from departures from LTE at deep atmospheric layers. The weak collisional coupling can be explained using the fact that the transitions of this level with levels of similar energy require two-electron (de)excitation.

\item The LTE profile of the 6162~\AA\ line (and other strong lines) does not reproduce the core and the wings of the line observed in the Sun. Typically, in the Sun, lines behave like the 6439~\AA\ line, profile fitting does not give a good match to the core in LTE, and as a result the LTE abundance increases and the macro-turbulent velocity decreases in order to fit the observations.   
 
\item In the three stars studied here, the \cai/\caii\ abundance-ratio discrepancy found in LTE is reduced with our NLTE modelling, improving the agreement between the derived \cai\ and \caii\ abundances. 
 
\item The CLV of the solar calcium lines is partially but not perfectly reproduced. For the lines studied here we suspect that the differences are related to the presence of inhomogeneities in the atmosphere. Strong lines such as the \caii\ triplet require a chromosphere due to the fact that their cores form at layers where the temperature inversion of the upper atmosphere cannot be ignored. 
 
\item Within the limitations of our modelling (1D, trace-element NLTE) we arrived at a solar calcium abundance that is consistent among lines over a large range of the solar spectrum and various formation depths; and also between neutral and singly ionised Ca. Nevertheless, it is important to note that our derived NLTE calcium abundance is about 0.05~dex lower than the meteoritic value (which is well within our uncertainties). 
 
\item In the three cases studied here, NLTE line profiles fit the observations better with respect to the LTE ones for the majority of the lines used here. When all lines are fit at once, Arcturus gives slightly better results in LTE in the visual region of the spectrum, but we find a negligible abundance correction \acorr=0.0. 
\end{itemize}

\noindent For these reasons, we believe that the present calculations are a step forward towards achieving more-accurate determinations of the calculated calcium abundances for late-type stars.

\begin{acknowledgements} 
K.L. acknowledge funds from the Alexander von Humboldt Foundation in the framework of the Sofja Kovalevskaja Award endowed by the Federal Ministry of Education and Research as well as funds from the Swedish Research Council (Grant nr. 2015-00415\_3) and Marie Sklodowska Curie Actions (Cofund Project INCA 600398). P.S.B. received financial support from the Swedish Research Council and the project grant ``The New Milky Way''  from the Knut and Alice Wallenberg Foundation . This work has made use of the VALD database, operated at Uppsala University, the Institute of Astronomy RAS in Moscow, and the University of Vienna. 
\end{acknowledgements}

\bibliographystyle{aa} 
\bibliography{papers,calcium}

\begin{appendix} 
 
\section {Table with line data}  
 
\input{bb_data_table.tex} 
 
\end{appendix}

\end{document}

%% file: bb_data_table.tex

\onecolumn
\centering
\longtab{
\begin{longtable}{rr@{$-$}lrcccccl}
\caption{ \label{tab:lines} Description of the transitions used in this work for comparison with observations. Columns 1 and 2 show wavelength and the levels involved, oscillator strength ($\log gf$) and source reference are in columns 3 and 4. van der Waals line broadening parameters and its reference source are in columns 5 and 6. Columns 7 and 8 show the Stark broadening parameter and its source reference. When available, Van der Waals broadening was calculated using the ABO theory with the broadening cross section $\sigma$ and the velocity parameter $\alpha$ shown in this table. The last column indicates the starts for which each line was used: `p' for Procyon, `s' for the Sun and `a' for Arcturus. References: K07 \citep{K07}, K10 \citep{K10}, ABO \citep{1995MNRAS.276..859A,1997MNRAS.290..102B,1998PASA...15..336B,1998MNRAS.296.1057B}, ORK-N \citep{1959ApJ...130..688O}, Y\&D \citep{YU2018263}, SG \citep{SG}, S \citep{S}, SN \citep{SN}, SR \citep{SR}, Sm \citep{Sm}, DIKH \citep{DIKH}, TB \citep{TB}, T \citep{T}.} \\
 \hline\hline
   \multicolumn{1}{c}{$\lambda$}  & \multicolumn{2}{c}{Transition}   & $\log{gf}$ & ref &  {$\log\Gamma_6$}                & ref & $\log{\Gamma_4/N_e}$  & ref  &  \\ 
          \multicolumn{1}{c}{[\AA]}   &    \multicolumn{2}{c}{ }   &                 &      & {[rad s$^{-1}$~cm$^3$]}         &      &   [rad s$^{-1}$~cm$^3$]  &  &   \\
                                                    &       \multicolumn{2}{c}{ }  &                 &      &$\sigma\small{[a.u.]}$\quad $\alpha$ &      &                                         &  & \\\hline
\endfirsthead
\caption{continued.}\\
\hline\hline
\multicolumn{1}{c}{$\lambda$} & \multicolumn{2}{c}{Transition}   & $\log{gf}$ & ref &  {$\log\Gamma_6$}                 & ref  & $\log{\Gamma_4/N_e}$ & ref  &  \\ 
          \multicolumn{1}{c}{[\AA]}   &    \multicolumn{2}{c}{ }   &                 &      & {[rad s$^{-1}$~cm$^3$]}         &      &   [rad s$^{-1}$~cm$^3$]  &  &   \\
                                                    &       \multicolumn{2}{c}{ }  &                 &      &$\sigma\small{[a.u.]}$\quad $\alpha$ &      &                                         &   & \\\hline
\endhead
\hline
\endfoot
       \multicolumn{10}{c}{\cai} \\\hline
 3\,875.80   & $ 4s.6f~^3\rm{F}^o_{3.0}$  & $ 3d.4s~^3\rm{D}_{3.0}$   &  -1.854 & ORK-N &    -7.467        &  K07  &  -3.687 & K07   &s   \\
 4\,094.91   & $ 4s.5f~^3\rm{F}^o_{3.0}$  & $ 3d.4s~^3\rm{D}_{2.0}$   &  -0.677 & ORK-N &    -7.441        &  K07  &  -3.981 & K07   &p   \\
 4\,108.53   & $ 4s.6f~^1\rm{F}^o_{3.0}$  & $ 3d.4s~^1\rm{D}_{2.0}$   &  -0.974 &  Y\&D &    -6.990        & K07   &  -2.860 & K07   &s   \\
 4\,226.72   & $ 4s.4p~^1\rm{P}^o_{1.0}$  & $ 4s.4s~^1\rm{S}_{0.0}$   &   0.244 &    SG & 371\quad 0.238   &  ABO  &  -6.031 &   SG  &p   \\ 
 4\,240.45   & $ 4s.7p~^1\rm{P}^o_{1.0}$  & $ 3d.4s~^1\rm{D}_{2.0}$   &  -1.673 &  Y\&D &    -7.030        & K07   &  -4.170 & K07   &s   \\
 4\,283.01   & $ 4p.4p~^3\rm{P}_{2.0}$    & $ 4s.4p~^3\rm{P}^o_{1.0}$ &  -0.423 &  Y\&D &    -7.720        & K07   &  -5.840 & K07   &ps  \\
 4\,298.99   & $ 4p.4p~^3\rm{P}_{1.0}$    & $ 4s.4p~^3\rm{P}^o_{1.0}$ &  -0.414 &  Y\&D &    -7.720        & K07   &  -5.760 & K07   &ps  \\
 4\,302.52   & $ 4p.4p~^3\rm{P}_{2.0}$    & $ 4s.4p~^3\rm{P}^o_{2.0}$ &   0.292 &     S &    -7.804        &  S    &  -5.997 &   S   &ps  \\
 4\,318.65   & $ 4p.4p~^3\rm{P}_{1.0}$    & $ 4s.4p~^3\rm{P}^o_{2.0}$ &  -0.222 & ORK-N &    -7.720        & K07   &  -5.760 & K07   &psa \\
 4\,355.07   & $ 4s.5f~^1\rm{F}^o_{3.0}$  & $ 3d.4s~^1\rm{D}_{2.0}$   &  -0.572 &  Y\&D &    -7.130        & K07   &  -3.550 & K07   &psa \\
 4\,425.43   & $ 4s.4d~^3\rm{D}_{1.0}$    & $ 4s.4p~^3\rm{P}^o_{0.0}$ &  -0.358 &  SN   &   946\quad 0.274 &  ABO  &  -5.610 &  SN   &psa \\
 4\,433.77   & $ 4s.7d~^1\rm{D}_{2.0}$    & $ 4s.4p~^1\rm{P}^o_{1.0}$ &  -2.562 & K07   &    -6.960        &  K07  &  -2.860 & K07   &p   \\
 4\,434.95   & $ 4s.4d~^3\rm{D}_{2.0}$    & $ 4s.4p~^3\rm{P}^o_{1.0}$ &  -0.279 &  Y\&D &   948\quad 0.274 &   ABO &  -5.602 & K07   &psa \\
 4\,435.67   & $ 4s.4d~^3\rm{D}_{1.0}$    & $ 4s.4p~^3\rm{P}^o_{1.0}$ &  -0.545 &  Y\&D &   947\quad 0.274 &   ABO &  -5.610 & K07   &psa \\
 4\,454.77   & $ 4s.4d~^3\rm{D}_{3.0}$    & $ 4s.4p~^3\rm{P}^o_{2.0}$ &   0.064 &  Y\&D &   949\quad 0.274 &   ABO &  -5.596 & K07   &psa \\
 4\,455.88   & $ 4s.4d~^3\rm{D}_{2.0}$    & $ 4s.4p~^3\rm{P}^o_{2.0}$ &  -0.518 &    SN &   948\quad 0.274 &   ABO &  -5.602 &  SN   &s   \\
 4\,456.61   & $ 4s.4d~^3\rm{D}_{1.0}$    & $ 4s.4p~^3\rm{P}^o_{2.0}$ &  -1.519 &  Y\&D &   947\quad 0.274 &   ABO &  -5.609 & K07   &psa \\
 4\,506.62   & $ 4s.6p~^3\rm{P}^o_{1.0}$  & $ 3d.4s~^3\rm{D}_{1.0}$   &  -2.357 &  Y\&D &    -7.230        & K07   &  -3.970 & K07   &s   \\
 4\,512.26   & $ 4s.6p~^3\rm{P}^o_{2.0}$  & $ 3d.4s~^3\rm{D}_{3.0}$   &  -1.900 &    SR &    -7.264        &  SR   &  -4.149 &  SR   &s   \\
 4\,581.36   & $ 4s.4f~^3\rm{F}^o_{10.0}$ & $ 3d.4s~^3\rm{D}_{2.0}$   &  -0.459 & ORK-N &    -7.631        & K07   &  -4.841 & K07   &s   \\
 4\,585.93   & $ 4s.4f~^3\rm{F}^o_{10.0}$ & $ 3d.4s~^3\rm{D}_{3.0}$   &  -0.313 & ORK-N &    -7.807        & K07   &  -5.017 & K07   &psa \\
 4\,685.26   & $ 4s.6d~^0\rm{D}_{2.0}$    & $ 4s.4p~^1\rm{P}^o_{1.0}$ &  -0.879 &     S &    -7.147        &   S   &  -4.503 &  S    &psa \\
 4\,847.31   & $ 4s.7s~^1\rm{S}_{0.0}$    & $ 4s.4p~^1\rm{P}^o_{1.0}$ &  -1.400 &     S &    -7.138        &   S   &  -4.161 &   S   &psa \\
 5\,041.61   & $ 4s.6p~^1\rm{P}^o_{1.0}$  & $ 3d.4s~^1\rm{D}_{2.0}$   &  -0.471 &    SR &    -7.308        &  SR   &  -4.294 &  SR   &s   \\
 5\,188.84   & $ 4s.5d~^1\rm{D}_{2.0}$    & $ 4s.4p~^1\rm{P}^o_{1.0}$ &  -0.075 &     S &    -7.315        &   S   &  -4.859 &  S    &sa  \\
 5\,260.38   & $ 3d.4p~^3\rm{P}^o_{2.0}$  & $ 3d.4s~^3\rm{D}_{1.0}$   &  -1.719 & SR    &   421\quad 0.260 &   ABO &  -5.756 &  SR   &psa \\
 5\,261.70   & $ 3d.4p~^3\rm{P}^o_{1.0}$  & $ 3d.4s~^3\rm{D}_{1.0}$   &  -0.579 & SR    &   421\quad 0.260 &   ABO &  -5.756 &  SR   &psa \\
 5\,262.24   & $ 3d.4p~^3\rm{P}^o_{0.0}$  & $ 3d.4s~^3\rm{D}_{1.0}$   &  -0.604 & ORK-N &   421\quad 0.261 &   ABO &  -5.756 & K07   &s   \\
 5\,265.55   & $ 3d.4p~^3\rm{P}^o_{1.0}$  & $ 3d.4s~^3\rm{D}_{2.0}$   &  -0.113 & SR    &   421\quad 0.260 &   ABO &  -5.755 &  SR   &ps  \\
 5\,349.46   & $ 3d.4p~^1\rm{F}^o_{3.0}$  & $ 3d.4s~^1\rm{D}_{2.0}$   &  -0.310 & SR+Sm &    -7.652        & SR+Sm &  -5.894 & SR+Sm &psa \\
 5\,512.97   & $ 4p.4p~^1\rm{S}_{0.0}$    & $ 4s.4p~^1\rm{P}^o_{1.0}$ &  -0.464 & S+Sm  &    -7.316        & S+Sm  &  -4.053 & S+Sm  &psa \\
 5\,581.96   & $ 3d.4p~^3\rm{D}^o_{3.0}$  & $ 3d.4s~^3\rm{D}_{2.0}$   &  -0.555 & SR    &   400\quad 0.282 &   ABO &  -6.072 &   SR  &psa \\
 5\,588.74   & $ 3d.4p~^3\rm{D}^o_{3.0}$  & $ 3d.4s~^3\rm{D}_{3.0}$   &   0.358 & SR    &   400\quad 0.282 &   ABO &  -6.072 & SR    &psa \\
 5\,590.11   & $ 3d.4p~^3\rm{D}^o_{2.0}$  & $ 3d.4s~^3\rm{D}_{1.0}$   &  -0.571 & SR    &   399\quad 0.282 &   ABO &  -6.071 & SR    &psa \\
 5\,594.46   & $ 3d.4p~^3\rm{D}^o_{2.0}$  & $ 3d.4s~^3\rm{D}_{2.0}$   &   0.097 & SR    &   399\quad 0.282 &   ABO &  -6.072 & SR    &psa \\
 5\,598.47   & $ 3d.4p~^3\rm{D}^o_{1.0}$  & $ 3d.4s~^3\rm{D}_{1.0}$   &  -0.087 & SR    &   398\quad 0.282 &   ABO &  -6.071 & SR    &psa \\
 5\,601.27   & $ 3d.4p~^3\rm{D}^o_{2.0}$  & $ 3d.4s~^3\rm{D}_{3.0}$   &  -0.523 & SR    &   399\quad 0.282 &   ABO &  -6.072 & SR    &psa \\
 5\,857.44   & $ 4p.4p~^1\rm{D}_{2.0}$    & $ 4s.4p~^1\rm{P}^o_{1.0}$ &  -0.051 &  Y\&D &  1438\quad 0.311 &   ABO &  -5.424 & K07   &psa \\
 5\,867.56   & $ 4s.6s~^1\rm{S}_{0.0}$    & $ 4s.4p~^1\rm{P}^o_{1.0}$ &  -1.570 &     S &    -7.460        &    S  &  -4.705 & S     &psa \\
 6\,102.71   & $ 4s.5s~^3\rm{S}_{1.0}$    & $ 4s.4p~^3\rm{P}^o_{0.0}$ &  -0.793 &   SN  &   876\quad 0.233 &   ABO &  -5.320 & SN    &psa \\
 6\,122.21   & $ 4s.5s~^3\rm{S}_{1.0}$    & $ 4s.4p~^3\rm{P}^o_{1.0}$ &  -0.332 &  Y\&D &   876\quad 0.234 &   ABO &  -5.320 & K07   &psa \\
 6\,161.29   & $ 4s.5p~^3\rm{P}^o_{2.0}$  & $ 3d.4s~^3\rm{D}_{2.0}$   &  -1.266 & SR    &   978\quad 0.257 &   ABO &  -4.994 & SR    &psa \\
 6\,162.17   & $ 4s.5s~^3\rm{S}_{1.0}$    & $ 4s.4p~^3\rm{P}^o_{2.0}$ &  -0.090 &    SN &   876\quad 0.234 &   ABO &  -5.320 & SN    &psa \\
 6\,163.75   & $ 4s.5p~^3\rm{P}^o_{1.0}$  & $ 3d.4s~^3\rm{D}_{1.0}$   &  -1.286 & SR    &   976\quad 0.257 &   ABO &  -4.998 & SR    &psa \\
 6\,166.44   & $ 4s.5p~^3\rm{P}^o_{0.0}$  & $ 3d.4s~^3\rm{D}_{1.0}$   &  -1.142 & SR    &   976\quad 0.257 &   ABO &  -4.999 &  SR   &psa \\
 6\,169.04   & $ 4s.5p~^3\rm{P}^o_{1.0}$  & $ 3d.4s~^3\rm{D}_{2.0}$   &  -0.797 & SR    &   976\quad 0.257 &   ABO &  -4.997 & SR    &psa \\
 6\,169.56   & $ 4s.5p~^3\rm{P}^o_{2.0}$  & $ 3d.4s~^3\rm{D}_{3.0}$   &  -0.478 & SR    &   978\quad 0.257 &   ABO &  -4.994 & SR    &psa \\
 6\,439.06   & $ 3d.4p~^3\rm{F}^o_{4.0}$  & $ 3d.4s~^3\rm{D}_{3.0}$   &   0.224 &  Y\&D &   366\quad 0.242 &   ABO &  -6.072 & K07   &psa \\
 6\,449.80   & $ 3d.4p~^1\rm{D}^o_{2.0}$  & $ 3d.4s~^3\rm{D}_{1.0}$   &  -0.502 & SR    &   365\quad 0.241 &   ABO &  -6.071 & SR    &psa \\
 6\,455.59   & $ 3d.4p~^1\rm{D}^o_{2.0}$  & $ 3d.4s~^3\rm{D}_{2.0}$   &  -1.340 &  S    &   365\quad 0.241 &   ABO &  -6.072 &  S    &psa \\
 6\,462.56   & $ 3d.4p~^3\rm{F}^o_{3.0}$  & $ 3d.4s~^3\rm{D}_{2.0}$   &   0.026 & SR+Sm &   365\quad 0.241 &   ABO &  -6.072 & SR+Sm &s   \\  
 6\,464.67   & $ 3d.4p~^3\rm{F}^o_{3.0}$  & $ 3d.4s~^1\rm{D}_{2.0}$   &  -2.103 & Y\&D  &    -7.690        &  K07  &  -5.810 & K07   &pa  \\  
 6\,471.65   & $ 3d.4p~^3\rm{F}^o_{3.0}$  & $ 3d.4s~^3\rm{D}_{3.0}$   &  -0.686 & SR+Sm &   365\quad 0.241 &   ABO &  -6.072 & SR+Sm &psa \\
 6\,493.78   & $ 3d.4p~^3\rm{F}^o_{2.0}$  & $ 3d.4s~^3\rm{D}_{1.0}$   &  -0.109 & SR+Sm &   364\quad 0.239 &   ABO &  -6.071 & SR+Sm &psa \\
 6\,499.65   & $ 3d.4p~^3\rm{F}^o_{2.0}$  & $ 3d.4s~^3\rm{D}_{2.0}$   &  -0.818 & SR+Sm &   364\quad 0.239 &   ABO &  -6.072 & SR+Sm &psa \\
 6\,508.84   & $ 3d.4p~^3\rm{F}^o_{2.0}$  & $ 3d.4s~^3\rm{D}_{3.0}$   &  -2.471 &  Y\&D &   364\quad 0.239 &   ABO &  -5.830 & K07   &sa  \\
 6\,572.77   & $ 4s.4p~^3\rm{P}^o_{1.0}$  & $ 4s.4s~^1\rm{S}_{0.0}$   &  -4.240 & DIKH  &   280\quad 0.242 &   ABO &  -6.031 & DIKH  &s   \\
 6\,717.68   & $ 4s.5p~^1\rm{P}^o_{1.0}$  & $ 3d.4s~^1\rm{D}_{2.0}$   &  -0.524 &    SR &   992\quad 0.255 &   ABO &  -4.895 & SR    &psa \\
 7\,148.14   & $ 3d.4p~^1\rm{D}^o_{2.0}$  & $ 3d.4s~^1\rm{D}_{2.0}$   &   0.137 &    SR &    -7.798        &  SR   &  -6.009 & SR    &psa \\
 7\,156.69   & $ 3d.4p~^3\rm{F}^o_{2.0}$  & $ 3d.4s~^1\rm{D}_{2.0}$   &  -3.713 &  Y\&D &    -7.710        &  K07   &  -5.830 & K07  &a   \\
 7\,202.20   & $ 3d.4p~^3\rm{F}^o_{2.0}$  & $ 3d.4s~^1\rm{D}_{2.0}$    &  -0.262 &    SR &    -7.798        &  SR   &  -6.009 & SR   &psa \\
 7\,326.13   & $ 4s.4d~^1\rm{D}_{2.0}$    & $ 4s.4p~^1\rm{P}^o_{1.0}$  &  -0.208 &     S &   845\quad 0.275 &   ABO &  -5.161 & S    &psa \\
 10\,275.38  & $ 4s.7d~^3\rm{D}_{7.0}$    & $ 4s.5p~^3\rm{P}^o_{2.0}$  &  -0.561 &  K07  &    -7.437        &   K07 &  -3.367 & K07  &s   \\
 10\,343.80  & $ 4s.5s~^1\rm{S}_{0.0}$    & $ 4s.4p~^1\rm{P}^o_{1.0}$  &   0.043 & Y\&D  &  1014\quad 0.221 &   ABO &  -5.060 & K07  &s   \\
 10\,833.37  & $ 3d.3d~^3\rm{P}_{2.0}$    & $ 3d.4p~^3\rm{P}^o_{1.0}$  &  -0.244 &  K07  &    -7.590        & K07   &  -4.710 & K07  &s   \\
 10\,838.96  & $ 3d.3d~^3\rm{P}_{2.0}$    & $ 3d.4p~^3\rm{P}^o_{2.0}$  &   0.238 &  K07  &    -7.590        & K07   &  -4.710 & K07  &s   \\
 10\,861.57  & $ 3d.3d~^3\rm{P}_{1.0}$    & $ 3d.4p~^3\rm{P}^o_{0.0}$  &  -0.343 &  K07  &    -7.590        & K07   &  -5.570 & K07  &s   \\
 10\,879.86  & $ 3d.3d~^3\rm{P}_{0.0}$    & $ 3d.4p~^3\rm{P}^o_{1.0}$  &  -0.357 &  K07  &    -7.590        & K07   &  -5.570 & K07  &s   \\
 12\,105.82  & $ 4s.6d~^1\rm{D}_{2.0}$    & $ 4s.5p~^1\rm{P}^o_{1.0}$  &  -0.478 &  Y\&D &    -7.090        & K07   &  -3.870 & K07  &s   \\
 12\,909.05  & $ 3d.3d~^3\rm{F}_{2.0}$    & $ 3d.4p~^3\rm{F}^o_{2.0}$  &  -0.452 &  Y\&D &    -7.710        & K07   &  -5.830 & K07  &s   \\
 13\,033.54  & $ 3d.3d~^3\rm{F}_{3.0}$    & $ 3d.4p~^3\rm{F}^o_{3.0}$  &  -0.193 &  Y\&D &    -7.710        & K07   &  -5.830 & K07  &s   \\
 13\,134.92  & $ 3d.3d~^3\rm{F}_{4.0}$    & $ 3d.4p~^3\rm{F}^o_{4.0}$  &  -0.043 &  Y\&D &    -7.710        & K07   &  -5.830 & K07  &s   \\
 13\,167.74  & $ 3d.3d~^3\rm{F}_{3.0}$    & $ 3d.4p~^3\rm{F}^o_{4.0}$  &  -1.113 &  Y\&D &    -7.710        & K07   &  -5.830 & K07  &s   \\
 15\,067.03  & $ 4s.np~^1\rm{P}^o_{1.0}$  & $ 4s.4d~^1\rm{D}_{2.0}$    &  -0.664 &  Y\&D &    -7.180        & K07   &  -4.110 & K07  &sa  \\
 16\,136.80  & $ 4s.5d~^3\rm{D}_{1.0}$    & $ 4s.5p~^3\rm{P}^o_{0.0}$  &  -0.973 &  Y\&D &    -7.250        & K07   &  -4.090 & K07  &a   \\
 16\,150.74  & $ 4s.5d~^3\rm{D}_{2.0}$    & $ 4s.5p~^3\rm{P}^o_{1.0}$  &  -0.369 &  Y\&D &    -7.250        & K07   &  -4.090 & K07  &sa  \\
 16\,155.22  & $ 4s.5d~^3\rm{D}_{1.0}$    & $ 4s.5p~^3\rm{P}^o_{1.0}$  &  -0.624 &  Y\&D &    -7.250        & K07   &  -4.090 & K07  &sa  \\
 16\,157.35  & $ 4s.5d~^1\rm{D}_{2.0}$    & $ 4s.5p~^1\rm{P}^o_{1.0}$  &  -0.411 &  Y\&D &    -7.260        & K07   &  -4.600 & K07  &sa  \\
 16\,197.06  & $ 4s.5d~^3\rm{D}_{3.0}$    & $ 4s.5p~^3\rm{P}^o_{2.0}$  &  -0.016 &  Y\&D &    -7.250        & K07   &  -4.080 & K07  &sa  \\
 19\,452.96  & $ 3d.4s~^3\rm{D}_{2.0}$    & $ 4s.4p~^3\rm{P}^o_{1.0}$  &  -0.958 &  Y\&D &    -7.770        & K07   &  -5.860 & K07  &s   \\
 19\,505.72  & $ 3d.4s~^3\rm{D}_{1.0}$    & $ 4s.4p~^3\rm{P}^o_{1.0}$  &  -1.215 &  Y\&D &    -7.770        & K07   &  -5.860 & K07  &s   \\
 19\,776.75  & $ 3d.4s~^3\rm{D}_{3.0}$    & $ 4s.4p~^3\rm{P}^o_{2.0}$  &  -0.617 &  Y\&D &    -7.770        & K07   &  -5.850 & K07  &s   \\
 19\,778.30  & $ 4p.4p~^1\rm{S}_{0.0}$    & $ 4s.5p~^1\rm{P}^o_{1.0}$  &  -0.649 &  Y\&D &    -7.440        & K07   &  -4.250 & K07  &sa  \\
 19\,815.00  & $ 4s.4f~^1\rm{F}^o_{3.0}$  & $ 4s.4d~^1\rm{D}_{2.0}$    &   0.273 &  Y\&D &  1363\quad 0.329 &   ABO &  -4.520 & K07  &sa  \\
 19\,853.07  & $ 4s.5p~^3\rm{P}^o_{2.0}$  & $ 4s.5s~^3\rm{S}_{1.0}$    &   0.105 &  Y\&D &  1131\quad 0.231 &   ABO &  -4.840 & K07  &sa  \\
 19\,862.17  & $ 3d.4s~^3\rm{D}_{2.0}$    & $ 4s.4p~^3\rm{P}^o_{2.0}$  &  -1.221 &  Y\&D &    -7.770        & K07   &  -5.850 & K07  &s   \\
 19\,917.17  & $ 3d.4s~^3\rm{D}_{1.0}$    & $ 4s.4p~^3\rm{P}^o_{2.0}$  &  -2.176 &  Y\&D &    -7.770        & K07   &  -5.850 & K07  &s   \\
 19\,933.71  & $ 4s.5p~^3\rm{P}^o_{1.0}$  & $ 4s.5s~^3\rm{S}_{1.0}$    &   0.101 &  Y\&D &  1130\quad 0.230 &   ABO &  -4.850 & K07  &sa  \\
 20\,962.39  & $ 4s.6p~^3\rm{P}^o_{2.0}$  & $ 4s.4d~^3\rm{D}_{3.0}$    &  -0.516 &  Y\&D &    -7.230        & K07   &  -3.960 & K07  &s   \\
 22\,624.33  & $ 4s.4f~^3\rm{F}^o_{10.0}$ & $ 4s.4d~^3\rm{D}_{2.0}$    &   0.074 &  Y\&D &  1259\quad 0.329 &   ABO &  -4.841 &  K07 &a   \\
 22\,821.04  & $ 4s.6p~^1\rm{P}^o_{1.0}$  & $ 4s.4d~^1\rm{D}_{2.0}$    &  -0.260 &  Y\&D &    -7.300        &  K07  &  -4.480 &  K07 &sa  \\\hline
      \multicolumn{10}{c}{\caii} \\\hline
  6\,456.87   & $ 6g~^2\rm{G}_{3.5}$   & $ 4f~^2\rm{F}_{3.5}$           &  -0.394 &  TB   &    -7.377        &  TB   &  -4.311 & TB   &ps \\
  8\,201.72   & $ 5d~^2\rm{D}_{1.5}$   & $ 5p~^2\rm{P}^o_{0.5}$         &   0.343 &  K10  &    -7.430        &  K    &  -4.980 & K10  &ps \\
  8\,248.78   & $ 5d~^2\rm{D}_{2.5}$   & $ 5p~^2\rm{P}^o_{1.5}$         &   0.556 &  TB   &    -7.412        &  TB   &  -4.600 & TB   &psa\\
  8\,254.71   & $ 5d~^2\rm{D}_{1.5}$   & $ 5p~^2\rm{P}^o_{1.5}$         &  -0.398 &  TB   &    -7.412        &  TB   &  -4.600 & TB   &ps \\
  8\,498.02   & $ 4p~^2\rm{P}^o_{1.5}$ & $ 3d~^2\rm{D}_{1.5}$           &  -1.416 &   T   &   291\quad 0.275 &   ABO &  -5.700 & T    &psa\\
  8\,542.08   & $ 4p~^2\rm{P}^o_{1.5}$ & $ 3d~^2\rm{D}_{2.5}$           &  -0.463 &   T   &   291\quad 0.275 &   ABO &  -5.700 & T    &psa\\
  8\,662.13   & $ 4p~^2\rm{P}^o_{0.5}$ & $ 3d~^2\rm{D}_{1.5}$           &  -0.723 &   T   &   291\quad 0.275 &   ABO &  -5.700 & T    &psa\\
  8\,912.06   & $ 4f~^2\rm{F}^o_{6.5}$ & $ 4d~^2\rm{D}_{1.5}$           &   0.637 &  TB   &    -7.512        &  TB   &  -5.100 & TB   &psa\\
  9\,854.75   & $ 6s~^2\rm{S}_{0.5}$   & $ 5p~^2\rm{P}^o_{0.5}$         &  -0.205 &  K10  &    -7.430        &  K    &  -5.010 & K10  &s  \\
  9\,890.62   & $ 5g~^2\rm{G}_{8.5}$   & $ 4f~^2\rm{F}^o_{6.5}$         &   1.268 &  TB   &    -7.607        &  TB   &  -4.541 & TB   &s  \\
  9\,931.37   & $ 6s~^2\rm{S}_{0.5}$   & $ 5p~^2\rm{P}^o_{1.5}$         &   0.092 &  K10  &    -7.430        &  K10  &  -5.010 & K10  &s  \\
  11\,838.99  & $ 5p~^2\rm{P}^o_{1.5}$ & $ 5s~^2\rm{S}_{0.5}$           &   0.290 & ATP-N &    -7.560        &  K10  &  -5.530 & K10  &s  \\
  11\,949.73  & $ 5p~^2\rm{P}^o_{0.5}$ & $ 5s~^2\rm{S}_{0.5}$           &  -0.010 & ATP-N &    -7.560        &  K10  &  -5.530 & K10  &s  \\
  16\,654.99  & $ 6d~^2\rm{D}_{4.5}$   & $ 6p~^2\rm{P}^o_{1.5}$         &   0.688 &  K10  &    -7.581        &  K10  &  -4.881 & K10  &s  \\
  21\,428.89  & $ 5d~^2\rm{D}_{1.5}$   & $ 4f~^2\rm{F}^o_{6.5}$         &   0.389 &  K10  &    -7.430        &  K10  &  -4.980 & K10  &s  \\       
\end{longtable}  
}